\input harvmac
\input amssym.def
\input amssym
\def\ds{\displaystyle}
\def\YBE{Yang-Baxter equation}
\def\LO{${\bf L}$-operator}
\def\RM{$R$-matrix}

\def\veps{\varepsilon}

\def\d{\delta}
\def\a{\alpha}
\def\b{\beta}
\def\l{\lambda}

\def\s{\sigma}

\def\lra{\longrightarrow}
\def\ra{\rightarrow}

\def\hf{{1\over2}}
 
\def\ds{\displaystyle}
\def\Tr{{\rm Tr}}
\def\frac#1#2{{#1\over#2}}

\input labeldefs.tmp
\writedefs

\lref\BLZ{Bazhanov, V.V., Lukyanov, S.L. and Zamolodchikov, A.B.:
Integrable structure of conformal field theory,
quantum KdV theory and
thermodynamic Bethe Ansatz.
Commun. Math. Phys. {\bf 177}, 381-398 (1996)}

\lref\BLZZ{Bazhanov, V.V., Lukyanov, S.L. and Zamolodchikov, A.B.:
Integrable structure of conformal field theory II.
$Q$-operator and DDV equation.
Commun. Math. Phys. {\bf 190}, 247-278 (1997)}

\lref\Baxter{Baxter, R.J.: Exactly solved models
in Statistical Mechanics. London: Academic Press, 1982}

\lref\FaL{Fateev, V.A. and  Lukyanov, S.L.:
Poisson-Lie group and classical W-algebras.
Int. J. Mod. Phys. {\bf A7}, 853-876 (1992)\semi
Fateev, V.A. and  Lukyanov, S.L.: Vertex operators and
representations of quantum universal enveloping algebras.
Int. J. Mod. Phys. {\bf A7}, 1325-1359 (1992)}

\lref\BLZZZ{Bazhanov, V.V., Lukyanov, S.L. and Zamolodchikov, A.B.:
Integrable quantum field theories in finite volume: Excited state energies. 
Nucl. Phys. {\bf B489}, 487-531 (1997)} 

\lref\BPZ{Belavin, A.A., Polyakov, A.M. and
Zamolodchikov, A.B.:
Infinite conformal symmetry in two-dimensional
quantum field theory.
Nucl. Phys. {\bf B241}, 333-380 (1984)}

\lref\Fei{Feigin, B.L. and  Fuchs, D.B.  Representations
of the Virasoro algebra. In:
Faddeev, L.D., Mal'cev, A.A. (eds.) Topology.
Proceedings, Leningrad 1982.
Lect. Notes in Math.
{\bf 1060}.
Berlin, Heidelberg, New York: Springer 1984}

\lref\DotsFat{ Dotsenko, Vl.S. and Fateev, V.A.:
Conformal algebra and
multipoint correlation functions in 2d statistical models. Nucl. Phys.
{\bf B240}\ [{\bf FS12}], 312-348 (1984)\semi
Dotsenko, Vl.S. and  Fateev, V.A.: Four-point
correlation functions and the operator algebra in 2d conformal invariant
theories with central charge $c\le1$.
Nucl. Phys. {\bf B251}\ [{\bf FS13}] 691-734 (1985) }

\lref\jap{Sasaki, R. and Yamanaka, I.:
Virasoro algebra, vertex operators, quantum Sine-Gordon
and solvable Quantum Field theories. Adv. Stud. in Pure Math.
{\bf 16}, 271-296 (1988) }

\lref\Eguchi{Eguchi, T. and Yang, S.K.:
Deformation of conformal field theories and soliton
equations. Phys. Lett. {\bf B224}, 373-378  (1989)}

\lref\Frenk{Feigin, B. and Frenkel, E.:
Integrals of motion and quantum groups.
In:  ``Integrable Systems and Quantum Groups", Proceeding of C.I.M.E.
Summer School, Lect. Notes in Math., {\bf 1620}, 349-418 (1996).
(hep-th/9310022)}

\lref\FF{Feigin, B. and Frenkel, E.:
Free field resolutions in affine Toda field theories.
Phys. Lett. {\bf B276}, 79-86 (1992)}

\lref\FFprivate{Feigin, B. and Frenkel, E.: private communication, 1997.}

\lref\Kor{Korepin, V.E., Bogoliubov, N.M. and Izergin, A.G.:
Quantum inverse  scattering method and correlation functions.
Cambridge University Press, 1993}

\lref\RKSA{Kulish, P.P., Reshetikhin, N.Yu. and  Sklyanin, E.K.:
Yang-Baxter equation and representation theory. Lett. Math.
Phys. {\bf 5}, 393-403 (1981) }

\lref\HL{Hardy, G.H. and Littlewood, J.E.: Notes on the Theory of
Series (XXIV): A Curious Power Series. Proc. Camb. Phil. Soc. 
{\bf 42}, 85-90 (1946)}

\lref\driver{Driver, K.A.,  Libinsky, D.S., Petruska, G. and Sarnak,
P.: Irregular distribution of $\{n\beta\}$,
$n=1,2,3,\ldots$, quadrature of singular integrals and curious basic
hypergeometric series. Indag. Mathem., N.S. {\bf 2}(4), 469-481 (1991)}

\lref\FSLS{Fendley, P., Lesage, F.  and Saleur, H.:
Solving 1d plasmas and 2d boundary problems using Jack
polynomial and functional relations. J. Statist. Phys. {\bf 79}, 799--819
 (1995)}

\lref\FadS{Sklyanin, E.K., Takhtajan, L.A. and Faddeev, L.D.:
Quantum inverse scattering method I.
Teoret. Mat. Fiz. {\bf 40}, 194-219 (1979) (In Russian)
[English translation:
Theor. Math. Phys. {\bf 40}, 688 - 706 (1979)].} 

\lref\KT{
Khoroshkin, S. M. and Tolstoy, V. N.: The uniqueness theorem for the
universal $R$-matrix.: Lett. Math. 
Phys. {\bf 24}, 231-244 (1992)}

\lref\KST{
Khoroshkin, S. M., Stolin, A. A. and Tolstoy, V. N.: Generalized Gauss
decomposition of trigonometric $R$-matrices. Mod. Phys. Lett.
{\bf A10},   1375-1392 (1995)}

\lref\TK{
Tolstoy, V. N. and Khoroshkin, S. M.: Universal $R$-matrix for
quantized nontwisted affine Lie algebras.
Funktsional. Anal. i Prilozhen. {\bf 26}(1), 85-88   (1992)  (In Russian)
[English translation: Func. Anal. Appl. {\bf 26}, 69-71  (1992)]}

\lref\Dr{Drinfel'd, V.G.: Quantum Groups. In: Proceedings of
the International Congress of Mathematics. Berkeley 1986,
{\bf 1}, pp. 798-820. California: Acad. Press 1987}

\lref\Jim{Jimbo, M.: A q-difference analogue of
\ $U{\cal G}$\ and Yang-Baxter
equation. Lett. Math. Phys. {\bf 10}, 63-68 (1985)}

\lref\Baxn{Baxter, R.J.:
Eight-vertex model
in lattice statistics and
one-dimensional anisotropic
Heisenberg chain\semi
1. Some fundamental 
eigenvectors. Ann. Phys. (N.Y.) {\bf 76}, 1-24 (1973)\semi
2. Equivalence to
a generalized ice-type model. Ann. Phys. (N.Y.)
{\bf 76}, 25-47 (1973)\semi
3. Eigenvectors of the transfer matrix and Hamiltonian.
Ann. Phys. (N.Y.) {\bf 76}, 48-71 (1973)}

\lref\KR{Kirilov, A.N. and Reshetikhin N.Yu.:
Exact solution of the integrable $ XXZ$  Heisenberg model with
arbitrary spin. J. Phys. {\bf A20}, 1565-1585 (1987)}

\lref\Vol{Volkov, A.Yu.: Quantum Volterra model. Phys. Lett. {\bf
A167}, 345-355 (1992)}
\lref\Volk{Volkov, A.Yu.: Quantum lattice KdV equation. 
Lett. Math. Phys. {\bf 39}, 313--329 (1997)}

%%%%%%%%%%%%%%%%%%%%%%%%%%%%%%%%%%%%%%%%%%%%%%%%%%%%%%%%%%%%%%%%%%%%%%%%%%

\Title{\vbox{\baselineskip12pt\hbox{RU-98-14}
\hbox{hep-th/9805008}}}
{\vbox{\centerline
{Integrable Structure of Conformal Field Theory III.}
\vskip6pt\centerline{
The Yang-Baxter Relation}}}
\centerline{Vladimir V. Bazhanov$^1$,
Sergei L. Lukyanov$^{2}$}
\centerline{ 
and Alexander B. Zamolodchikov$^{2}$ }
\centerline{ }
\centerline{$^1$Department of 
Theoretical Physics and Center of Mathematics}
\centerline{and its Applications, IAS, Australian National University, }
\centerline{Canberra, ACT 0200, Australia}
\centerline{and}
\centerline{Saint Petersburg Branch of Steklov Mathematical Institute,}
\centerline{Fontanka 27, Saint Petersburg, 191011, Russia}
\centerline{ }
\centerline{$^2$Department of Physics and Astronomy,}
\centerline{Rutgers University, Piscataway, NJ 08855-049, USA}
\centerline{and}
\centerline{L.D. Landau Institute for Theoretical Physics,}
\centerline{Chernogolovka, 142432, Russia}

\Date{April, 98}
\vfil
\eject

\centerline{{\bf Abstract}}

In this paper we fill some gaps in the arguments of our previous papers
[1,2]. In particular, we give a proof that the ${\bf L}$ operators of 
Conformal Field Theory indeed satisfy the defining relations of 
the Yang-Baxter algebra. Among other results we  present a derivation 
of the functional 
relations satisfied by ${\bf T}$ and ${\bf Q}$ operators and a proof of 
the basic analyticity assumptions for these operators used in [1,2].

\vfil
\eject

\newsec{Introduction}

This paper is a sequel to our
works\ \refs{\BLZ,\BLZZ} where we have introduced the
families of operators ${\bf T}(\lambda)$ and ${\bf
Q}(\lambda)$ which act in a highest weight Virasoro
module and satisfy the
commutativity conditions
\eqn\TQcomm{
[{\bf T}(\lambda), {\bf T}(\lambda')] =  
[{\bf T}(\lambda), {\bf Q}(\lambda')] = [{\bf Q}(\lambda), {\bf
Q}(\lambda')] = 0\, . 
}
These operators are CFT analogs of Baxter's commuting transfer-matrices 
of integrable lattice theory\ \refs{\Baxn, \Baxter}. 
In the lattice theory the transfer-matrices
are typically constructed as follows. One first finds an $R$-matrix
which solves the Yang-Baxter equation 
\eqn\Ybe{
{\bf R}_{V V'}(\lambda)\, {\bf R}_{V V''}(\lambda\lambda')\, {\bf R}_{V' V''}
(\lambda')= {\bf R}_{V' V''}(\lambda')\, {\bf R}_{V V''}
(\lambda\lambda')\, {\bf R}_{V V'}(\lambda)\ .
}
Here ${\bf R}_{VV'},\, {\bf R}_{V V''},\, 
{\bf R}_{V' V''}$ act in the tensor product of the 
identical vector spaces $V$, $V'$ and $V''$. Then one introduces the 
$L$-operator 
\eqn\lop{
{\bf L}_V (\lambda) = {\bf R}_{V V_1}(\lambda){\bf R}_{V V_2}
(\lambda)\ldots {\bf R}_{V
V_N}(\lambda)\ ,}
which is considered as a matrix in 
$V$ whose elements are operators acting
in the tensor product
\eqn\hlat{
{\cal H}_N = \otimes_{i=1}^{N}\,  V_i\ ,}
where $N$ is the size of the lattice. The space \hlat\ is interpreted as
the space of states of the lattice theory. The operator \lop\ satisfies the
defining relations of the {\it Yang-Baxter algebra}
\eqn\yba{
{\bf R}_{V V'}(\lambda/\lambda')\, 
{\bf L}_{V}(\lambda){\bf L}_{V'}(\lambda')=
{\bf L}_{V'}(\lambda'){\bf L}_{V}(\lambda)\, 
{\bf R}_{V V'}(\lambda/\lambda')\ . 
}
It realizes, thereby, a representation of this 
algebra in the space of states of the lattice theory.
The ``transfer-matrix''
\eqn\tlat{
{\bf T}_V(\lambda) = {\rm Tr}_V \big[\, {\bf L}_{V}(\lambda)\, \big]\, :\ 
{\cal H}_N\to {\cal H}_N
}
satisfies the commutativity condition \TQcomm\  as a simple 
consequence of the defining relations \yba.

In many cases the integrable lattice theory defined through the 
transfer-matrix \tlat\ can be used as the starting point to construct
an integrable quantum field theory (QFT). If the lattice system has a
critical point one can define QFT by taking appropriate scaling limit
(which in particular involves the limit $N\to \infty$). Then the space of
states of QFT appears as a certain subspace in the limiting space
\hlat, ${\cal H}_{QFT} \subset {\cal H}_{N\to\infty}$. Although many
integrable QFT can be constructed and studied this way (and this
is essentially the way integrable QFT are obtained in Quantum Inverse
Scattering Method\ \refs{\FadS, \Kor} ), the alternative idea of constructing
representations of Yang-Baxter algebra directly in the space of states
${\cal H}_{QFT}$ of continuous QFT seems to be more attractive. This idea
was the motivation of our constructions in \refs{\BLZ,\BLZZ}. 

The natural starting point for implementing this idea is the Conformal
Field Theory (CFT) because the general structure of its space of states
${\cal H}_{CFT}$ is relatively well understood\ \BPZ. The space
${\cal H}_{CFT}$ can be decomposed as 
\eqn\Hcft{
{\cal H}_{CFT} = \oplus_{_{\Delta, {\overline \Delta}}}\,
 \big[\, {\cal V}_{_\Delta}
\otimes {\overline{\cal V}}_{_{\overline \Delta}}\, \big]\ ,
}
where ${\cal V}_{_\Delta}$ and ${\overline{\cal V}}_{_{\overline \Delta}}$ are 
irreducible highest weight
representations of ``left'' and ``right'' Virasoro algebras with the
highest weights $\Delta$ and\ ${{\overline \Delta}}$ respectively. 
The sum \Hcft\ may be finite (as
in the ``minimal models''), infinite or even continuous. In any case the
space \Hcft\ can be embedded into a direct product 
${\cal H}_{\rm chiral}\otimes
{\overline
{\cal H}}_{\rm chiral}$ of left and right ``chiral'' subspaces,
\eqn\Hc{
{\cal H}_{\rm chiral} = \oplus_{_\Delta} {\cal V}_{_\Delta}\,.
}
In \refs{\BLZ,\BLZZ} we introduced the operators ${\bf L}(\lambda)$
which realize 
particular representations of the Yang-Baxter algebra \yba\ in the 
space \Hc. The commuting operators \TQcomm\  was constructed in terms
of these operators ${\bf L}$. However, the proof that
these operators actually satisfy the defining relations \yba\ of the
Yang-Baxter algebra was not presented.
The main purpose of this paper is to fill this gap.

Here we remind some notations used 
in \refs{\BLZ,\BLZZ}. Let $\varphi(u)$ be a free chiral 
Bose field, i.e. the operator-valued function
\eqn\vars{\varphi(u) = iQ + i P u + \sum_{n\neq 0}
{{a_{-n}}\over {n}}\  
e^{inu}\ , }
where $ P,  Q$ and $ a_n\ ( n = \pm 1, \pm 2, \ldots)$\ are
operators which satisfy
the commutation relations of the Heisenberg
algebra
\eqn\osc{[ Q, P] = {i\over 2}\ \beta^2\, ; \qquad [ a_n ,  a_m] = 
{n\over 2}\ \beta^2\  \delta_{n+m, 0}\ .}
with real $\beta$.
The variable  $u$ is interpreted as a complex coordinate on $2D$ 
cylinder of a circumference $2\pi$. The field $\varphi (u)$ is a 
quasi-periodic function of $u$, i.e.
\eqn\perka{\varphi(u+2\pi) = \varphi(u) + 2\pi i P\ .}
Let ${\cal F}_p$ be the Fock space, i.e. the space
generated by a free action 
of the operators $a_n$ with $n < 0$ on the vacuum
vector $\mid p \rangle$ which satisfies
\eqn\kuyhg{\eqalign{
&a_n \mid p \rangle = 0\ , \quad {\rm  for} \quad n > 0\ ;\cr
& P\mid p \rangle = p\mid p \rangle\ .}} 
The space ${\cal F}_p$ supports a highest weight representation of the
Virasoro algebra generated by the operators
\eqn\tens{L_n = \int_{0}^{2\pi}{du\over{2\pi}}\ 
\Big[\, T(u) + {c\over 24}\, \Big] \, e^{inu} }
with the Virasoro central charge
\eqn\central{
c = 13-6\, \big(\beta^2+\beta^{-2}\big)\ 
}
and the highest weight 
\eqn\juyt{
\Delta =\Delta(p) \equiv \Big(\frac{p}{\beta}\Big)^2 +\frac{c-1}{24}\ .
}
Here $T(u)$ denotes the composite field 
\eqn\nuyt{- \beta^2\,  T(u) =
:\varphi' (u)^2: + (1 - \beta^2)\, \varphi''(u) + 
{{\beta^2}\over 24}  }
which is a periodic function, \ $T(u+2\pi)=T(u)$. 
The symbol \  $:\ \ :$\  denotes the standard normal ordering with
respect to the Fock vacuum \kuyhg.
It is well known that if the parameters $\beta$ and $p$ take generic
values this representation of the Virasoro algebra is
irreducible. For particular values of these 
parameters, when null-vectors appear in ${\cal F}_p$, the irreducible 
representation ${\cal V}_{\Delta(p)}$ is obtained from ${\cal F}_p$ by 
factoring out all the invariant subspaces. In what follows we will always
assume that all the invariant subspaces (if any) are factored out, and
identify the spaces ${\cal F}_p$ and ${\cal V}_{\Delta(p)}$.

The space
\eqn\mju{{\hat {\cal F}}_{p}=
\oplus_{k=-\infty}^{\infty}\,  {\cal F}_{p+k{\beta^2}} }
admits the action of the exponential fields 
\eqn\eip{V_{\pm}(u) = :e^{\pm 2 \varphi (u)}:\equiv
\exp\big(\pm 2 \sum_{n=1}^{\infty}{a_{-n}\over
n}e^{inu}\big)\,\exp\big(\pm 2  i\ (Q+Pu)\big)\,\exp\big(\mp 2 
\sum_{n=1}^{\infty}{a_{n}\over n}e^{-inu}\big)\ .}
The following relations are easily verified
from \vars-\perka\ 
\eqn\vcomm{\eqalign{
V_{\sigma_1}(u_1)\ V_{\sigma_2}(u_2)&=q^{2\s_1\s_2}\ 
V_{\sigma_2}(u_2)\ V_{\sigma_1}(u_1)\, ,\qquad u_1>u_2\, ,
\cr
P\  V_{\pm}(u)&=  V_{\pm}(u)\ (P\pm\beta^2)\ ,
\cr}}
where $\s_1,\s_2=\pm 1$. Moreover,
\eqn\vper{V_\pm(u+2\pi)=q^{-2}\,e^{\pm4\pi i P}\,V_\pm(u)\ .}

Any CFT possesses infinitely many local Integrals of 
Motion (IM) ${\bf I}_{2k-1}$ \refs{\jap, \Eguchi}
\eqn\loim{{\bf I}_{2k-1} = \int_{0}^{ 2\pi} {{du}\over {2\pi}}\ 
T_{2k}(u)\ ,\qquad k=1,2\ldots\ ,}
where $T_{2k}(u)$ are certain local fields, polynomials in $T(u)$ and 
its derivatives. For example
\eqn\locdens{\eqalign{T_2 (u) = T&(u)\, , \ \   T_4 (u) = :T^2 (u):\, , \  
\ \ \ldots\, , \cr 
&T_{2k}(u) = : T^k (u):\  + \ {\rm terms\  with\  the\
derivatives}\ . }}
Here $:\ \ :$ denote appropriately regularized operator products,
see\ \BLZ\ for details.
There exists infinitely many densities \locdens\  (one
for each integer\ $k$\ \refs{\Frenk,\FF}) such that all
IM\loim\  commute
\eqn\locomm{[\, {\bf I}_{2k-1}, {\bf I}_{2l-1}] = 0\ .}

Consider the following operator matrix\ \refs{\FaL, \BLZ}\foot{Note
that the discrete  analog of the operator ${\bf L}_{1\over2}(\l)$
has been used in \refs{\Vol,\Volk} in the context of the quantum lattice 
KdV equation.}
\eqn\Lj{{\bf L}_j (\lambda) =\pi_j\big[\ {\bf L} (\lambda)\ \big]\ ,}
\eqn\Lop{{\bf L} (\lambda) = e^{i \pi P H}\ 
 {\cal P}\exp\bigg\{\lambda
\int_{0}^{2\pi}du \big(\, V_-(u)\, q^{H\over 2}E + V_+(u)\, q^{-{H\over 2}}F\,
\big)\bigg\}\ ,}
where the exponential fields $V_\pm(u)$  are defined in \eip\
and $E, F$ and $H$ are the generating elements of the quantum universal
enveloping algebra  $U_q\big(sl(2)\big) $\ \RKSA,
\eqn\uals{[H,E]=2E\, ,
\qquad [H,F]=-2F\, , \qquad [E,F]={{q^{H}-q^{-H}}\over {q -
q^{-1}}}\ ,}
with
\eqn\qpar{q=e^{i\pi \beta^2}\ .}
The symbol $\pi_j$ in\ \Lj\ stands  for the $(2j+1)$
dimensional representation of $U_q\big(sl(2)\big)$, so that 
\Lj\  is in fact 
$(2j+1)\times (2j+1)$ matrix whose elements are the operators acting in
the space \mju. Following the conventional terminology, we will  
refer to this space as the ``quantum space''.
The expression \Lj\ contains the ordered exponential (the symbol ${\cal
P}$ denotes the path ordering) which can be defined in terms of the
power series in $\lambda$ as follows,
\eqn\troi{{\bf L}_j (\lambda)=
\pi_j \bigg [ e^{i\pi P H} \sum_{k=0}^{\infty}
\lambda^k \int_{2\pi\geq u_1 \geq u_2 \geq ... \geq
u_k\geq 0}K(u_1)K(u_2)...K(u_k)\ du_1 du_2 ... du_k\bigg ]\ ,}
where
\eqn\Kdef{
K(u)= V_-(u)\, q^{H\over 2}E+V_+(u)\, q^{-{H\over 2}}F\ . }
The integrals in\  \troi\  make perfect sense if 
\eqn\cent{-\infty < c < -2\ .}
For $-2 < c < 1$ the integrals \troi\ diverge and power series
expansion of \Lop\  
should be written down in terms
of contour integrals, as explained in \BLZZ\ (see also Appendix~C of
this paper).  

In Sect.2 we will show that the operator matrices \Lj\ satisfy the
relations \yba,
\eqn\yab{{\bf R}_{jj'}(\lambda\mu^{-1})\ 
\big({\bf L}_j(\lambda)\otimes 1\big)\ 
\big(1\otimes {\bf L}_{j'}(\mu)\big)=
\big(1\otimes {\bf L}_{j'}(\mu)\big) \ 
\big({\bf L}_{j}(\lambda)\otimes 1\big)\ {\bf R}_{jj'}
(\lambda \mu^{-1})\ ,}
where the matrix \ ${\bf R}_{jj'}(\lambda)$\ is the $R$-matrix
associated with the representations $\pi_j$, $\pi_j'$ of $U_q\big(sl(2)\big)$; 
in particular
\eqn\bxvc{\eqalign{{\bf R}_{{1\over 2} {1\over 2}}(\lambda)
=\pmatrix{q^{-1}\lambda-q \lambda^{-1}&
{}&{}&{}\cr
{}&\lambda-\lambda^{-1}&q^{-1}-q&{}\cr
{}&q^{-1}-q&\lambda-\lambda^{-1}&{}\cr
{}&{}&{}&q^{-1}\lambda-q \lambda^{-1}}\ .}}
coincides with the $R$-matrix of the six-vertex model.
With an appropriate normalization the matrix 
${\bf R}_{jj'}(\lambda)$ is a finite  Laurent polynomial in $\l$.   
Therefore, after multiplication by a simple power factor  
both sides of \yab\ can be expanded in infinite series in the 
variables $\l$ and $\mu$. 
We will prove that the relations \yab\ are valid 
to all orders of these expansions. In fact, we will construct more
general ${\bf L}$-operators which satisfy the Yang-Baxter relation
\yba\ with the universal $R$-matrix for the quantum Kac-Moody algebra 
$U_q\big(\widehat{sl}(2)\big)$. The equation \yab\ will follow then as
a particular case. 

\newsec{The Yang-Baxter relation}
The 
quantum Kac-Moody algebra ${\cal A}=U_q\big(\widehat{sl}(2)\big)$
is generated by elements $h_0, h_1$, $x_0$, $x_1$, $y_0$, $y_1$, subject
to the commutation relations
\eqn\KMsimple{
 \left[h_i,h_j\right]=0\, ,\qquad
\left[h_i, x_j\right]= -a_{ij} x_j\, ,\qquad 
\left[h_i, y_j\right]= a_{ij} y_j\, ,
}
\eqn\KMrel{
\left[y_i, x_j\right]=
\delta_{i j}\ {q^{h_i}-q^{-h_i}\over q-q^{-1}}\ ,}
and the Serre relations
\eqn\serre{\eqalign{
&x_i^3 x_j-\left[3\right]_q x_i^2 x_j x_i+
\left[3\right]_q x_i x_j x_i^2-x_j x_i^3=0\, ,\cr
&y_i^3 y_j-\left[3\right]_q y_i^2 y_j y_i+
\left[3\right]_q y_i y_j y_i^2-y_j y_i^3=0\,  .\cr
}}
Here the indices $i,j$ take two values $i,j=0,1$;\ $a_{ij}$ is the
Cartan matrix of the algebra  $U_q\big(\widehat{sl}(2)\big)$,
$$a_{ij}=\pmatrix{2&  -2\cr -2 &2}\ ,$$ and  
 $[n]_q=(q^n-q^{-n})/(q-q^{-1})$.
The sum
\eqn\cen{k=h_0+h_1 }
is a  central element in the algebra ${\cal A}$.
Usually the algebra\ ${\cal A}$ \ is 
supplemented by the grade operator\ $d$\
\eqn\iuy{[d,h_0]=[d,h_1]=[d,x_0]=[d,y_0]=0\, ,
\ \ \ \  \ \ [d,x_1]=x_1,\ [d,y_1]=-y_1\ .}
The algebra ${\cal A}=U_q\big(\widehat{sl}(2)\big)$ is a Hopf algebra with
the co-multiplication
$$
\delta :  \qquad  {\cal A} \lra {\cal A}\otimes {\cal A}
$$
defined as
\eqn\coproduct{\eqalign{
&\delta(x_i)=x_i\otimes 1+q^{-h_i}\otimes x_i\ ,\cr
&\delta(y_i)=y_i\otimes q^{h_i}+1\otimes y_i\ ,\cr
&\delta(h_i)=h_i\otimes 1+1\otimes h_i\ ,\cr
&\delta(d)=d\otimes 1+1\otimes d\ .\cr}}
where $i=0,1$.   
As usual we introduce 
\eqn\Delp{
\delta'=\sigma\circ\delta\, , \qquad \sigma\circ(a\otimes b)=b\otimes a
\qquad  (\, \forall a,b\in {\cal A}\, )\ .
}
Define also two Borel subalgebras ${\cal B}_-\subset {\cal A}$
and  ${\cal B}_+\subset {\cal A}$ generated by $d,\, h_{0,1},\, 
x_0,\, x_1$ and
$d,\, h_{0,1},\, y_0,\, y_1$  respectively. 
There exists a unique element \refs{\Dr,\KT} 
\eqn\imbed{
{\cal R}\in {\cal B}_+\otimes {\cal B}_-\ ,
}
 satisfying the following relations 
\eqn\defineR{\eqalign{
&\delta'(a)\ {\cal R}={\cal R}\ \delta(a)
\qquad (\forall\ a\in {\cal A})\, ,\cr
&(\delta\otimes 1)\, {\cal R}={\cal R}^{13}\, {\cal R}^{23}\, ,\cr
&(1\otimes \delta)\, {\cal R}={\cal R}^{13}\, {\cal R}^{12}\, ,\cr
}}\bigskip\noindent
where ${\cal R}^{12},\, {\cal R}^{13},\, {\cal R}^{23}\in
{\cal A}\otimes{\cal A}\otimes {\cal A}$ and
${\cal R}^{12}={\cal R}\otimes 1$, ${\cal R}^{23}=1\otimes {\cal R}$,
${\cal R}^{13}=(\sigma\otimes 1)\, {\cal R}^{23}$.
The element ${\cal R}$ is called the universal $R$-matrix. 
It satisfies the \YBE\ 
\eqn\ybe{
{\cal R}^{12}{\cal R}^{13}{\cal R}^{23}={\cal R}^{23}{\cal
R}^{13}{\cal R}^{12}\ ,}
which is a simple corollary of the definitions \defineR . 
The universal 
$R$-matrix is understood as a formal series in generators in ${\cal
B}_+ \otimes {\cal B}_-$. Its dependence on the Cartan elements 
can be isolated as a simple factor. It will be convenient
to  introduce the ``reduced" universal $R$-matrix 
\eqn\rseries{
\overline{{\cal R}}=q^{-(h_0\otimes h_0) 
/2+k\otimes d+d\otimes k}\  {\cal R} = 
(\hbox{series in\ } y_0,y_1,x_0,x_1)\ ,
}
where 
$y_i\in {\cal B}_+\otimes 1$, $x_i\in 1\otimes{\cal B}_-$\  $(i=0,1)$.
There exists an ``explicit'' expression for the universal $R$-matrix
\refs{\TK,\KST} which, in general case,  provides 
an algorithmic procedure for the computation 
of this series order by order. Using these results or directly from
the definitions \imbed\ and \defineR\ 
one can calculate  the first few terms in \rseries\ 
\eqn\rexpan{\eqalign{
\overline{{\cal R}}=
1+(q-q^{-1})&(y_0\otimes x_0+y_1\otimes x_1)+
{q-q^{-1}\over \left[2\right]_q}\bigg\{(q^2-1)( y_0^2\otimes x_0^2+
y_1^2\otimes x_1^2)+\cr &
y_0 y_1\otimes( x_1 x_0-q^{-2} x_0 x_1)+
y_1 y_0\otimes (x_0 x_1-q^{-2} x_1 x_0)
\bigg\}+\dots \ .\cr}}
The higher terms  soon become very complicated and their
general form is unknown. 
This complexity  should not be surprising, since the universal $R$-matrix 
contains infinitely many nontrivial 
solutions of the \YBE\ associated with 
$U_q\big(\widehat{sl}(2)\big)$. 
A few more terms of the expansion \rexpan\ are given
in the Appendix~A.

We are now ready to prove the Yang-Baxter equation \yab\ and its
generalizations. Consider 
the following  operator
\eqn\Lopg{
{\cal L}=e^{i\pi Ph}{\cal P} \exp\Big( \int_0^{2\pi} {\cal K}(u) du \Big)\ ,
}
where 
\eqn\calK{
{\cal K}(u)=V_-(u)\, y_0\ +V_+(u)\, y_1\  .}
Here $h=h_0=-h_1,\,  y_0,\,  y_1$ are the generators of  
the Borel subalgebra\ ${\cal B}_+$ and the ${\cal P}$-exponent is
defined  as the series of the ordered integrals of ${\cal K}(u)$, 
similarly to \troi. Notice that we assumed here that the central charge
$k$ is zero; considering this case is sufficient for our goals. 
The operator \Lopg\ is an element of the algebra 
${\cal B}_+$ whose coefficients are operators acting in the quantum 
space \mju.  It is more general than the one in \Lop\ and reduces to
the latter for a particular  representation of ${\cal B}_+$ (see below).
Consider now two operators \Lopg 
\eqn\Ltwo{
{\cal L}\otimes1\in {\cal B}_+\otimes 1,\qquad 1\otimes{\cal L}\in
1\otimes{\cal B}_+ }
belonging to the different factors of the direct product
${\cal B}_+\otimes{\cal B}_+$.
Using \vcomm\ 
for the product of these operators 
one obtains  
\eqn\Lpro{
(\,{\cal L}\otimes1\,) \,(\,1\otimes{\cal L}\,)=
e^{i\pi P\delta(h) }\ {\cal P} \exp\Big( \int_0^{2\pi} 
{\cal K}_1(u) du \Big)\
{\cal P} \exp\Big( \int_0^{2\pi} {\cal K}_2(u) du \Big)\ ,}
where 
\eqn\delh{
\delta(h)=h\otimes1+1\otimes  h\ ,
}
and
\eqn\Kot{\eqalign{
{\cal K}_1(u)=&V_-(u)\, (y_0\otimes
q^h)\ +V_+(u)\, (y_1\otimes q^{-h})\ ,
\cr
{\cal K}_2(u)=&V_-(u)\, (1\otimes
y_0)\ +V_+(u)\, (1\otimes y_1)\ .
}}
Taking into account \vcomm\ and \KMsimple\ it is easy to see that
\eqn\Kcomm{
[{\cal K}_1(u_1),{\cal K}_2(u_2)]=0\,  , \qquad u_1<u_2\ ,
}
therefore the product of the ${\cal P}$-exponents in \Lpro\ can be
rewritten as
\eqn\Lprod{\eqalign{
(\,{\cal L}\otimes1\,) \,(\,1\otimes{\cal L}\,)&=
e^{i\pi P\delta(h) }\ {\cal P} \exp\Big( \int_0^{2\pi} \big({\cal
K}_1(u)+{\cal K}_2(u)\big)  du \Big)\cr
&=e^{i\pi P\delta(h) }\ {\cal P} \exp\Big( \int_0^{2\pi} \big(
V_-(u)\,\delta(y_0)\,\,+V_+(u)\,\delta(y_1)\,
\big)  du \Big)\cr
&=\delta({\cal L})\ ,}}
where the co-multiplication $\delta$ is defined in \coproduct.
Similarly
\eqn\Ldorp{
(\,1\otimes{\cal L}\,)\,(\,{\cal L}\otimes1\,)=\delta'({\cal L})\ , 
}
with $\delta'$ defined in \Delp. Combining \Lprod\ and \Ldorp\ with the
first equations in \defineR\ one obtains the following \YBE\
\eqn\ybeL{
{\cal R}\,\,(\,{\cal L}\otimes1\,) \,(\,1\otimes{\cal L}\,)=
(\,1\otimes{\cal L}\,)\,(\,{\cal L}\otimes1\,) \,\, {\cal R}\ .
}
Obviously, this equation is more general than \yab.
To obtain the latter from \ybeL\ we only need to choose appropriate
representations in each factor of the direct product ${\cal A}\otimes
{\cal A}$ involved in \ybeL. Consider the  so-called 
{\it evaluation  homomorphism} $U_q\big(\widehat{sl}(2)\big)\lra U_q
\big(sl(2)\big)$
of the form
\eqn\evalr{\eqalign{
x_0 \ra \l^{-1}  F q^{-H/2}\, ,\qquad &y_0 \ra \l  q^{H/2} E\, ,\qquad
h_0\ra H\, , \cr 
x_1 \ra \l^{-1}E  q^{H/2}\, ,\qquad &y_1\ra \l   q^{-H/2}F\, ,\qquad 
h_1\ra -H\, ,\cr}
}
where $\l$ is a spectral parameter,
and $E,\, F,\, H$ are the generators 
of the  algebra $U_q\big(sl(2)\big)$, defined already in \uals.
One 
could easily check that with the map \evalr\ all the 
defining  relations \KMsimple, \KMrel\ and \serre\ of the algebra
${\cal A}=U_q\big(\widehat{sl}(2)\big)$ become simple corollaries
of \uals. 
For any representation $\pi$ of $U_q(sl(2))$ the formulae 
\evalr\ define a representation of the algebra ${\cal A}$
with zero central charge $k$,  which will
be denoted as $\pi(\l)$. In particular, the matrix representations of
${\cal A}$ corresponding to the $(2j+1)$-dimensional
representations $\pi_j$ of $U_q\big({sl}(2)\big)$ will be denoted
$\pi_j(\l)$. Let us now evaluate the \YBE\ \ybeL\ in the
representations $\pi_j(\l)$ and $\pi_{j'}(\mu)$ for  the first and
second factor of the direct product respectively.
For the ${\bf L}$-operators one has
\eqn\Lja{
\pi_j(\l)\big[{\cal L}\big]={\bf L}_j(\l)\, ,\qquad 
\pi_{j'}(\mu)\big[{\cal L}\big]={\bf L}_{j'}(\mu)\, ,
}
with ${\bf L}_j$ given by \Lj,
while for the $R$-matrix one obtains
\eqn\Rjj{
\big(\pi_j(\l)\otimes\pi_{j'}(\mu)\big)\big[{\cal R}\big]
 = \rho_{jj'}(\l/\mu)\ {\bf R}_{jj'}(\l/\mu)\ ,
}
where $\rho_{jj'}$ is a scalar factor and the  ${\bf
R}_{jj'}$ is the same as in \yab\ \Jim. This completes the proof of \yab.

We conclude this section with following observation
concerning the structure of the \LO\ \Lopg.
As one could expect the equation \ybeL\ is, in fact, a specialization of
the \YBE\ \ybe\ for the universal \RM. To demonstrate this it would be
sufficient 
to find an appropriate realization of the algebra ${\cal A}$ in the 
third factor of the product ${\cal A}\otimes{\cal A}
\otimes {\cal A}$ involved in \ybe, such that \ybe\ reduces to \ybeL.
A little inspection  shows that each side of
\ybe\  is an element of ${\cal B}_+\otimes{\cal A}
\otimes {\cal B}_-$ rather than an element of ${\cal A}\otimes{\cal A}
\otimes {\cal A}$. Therefore we do not need a realization 
of the full algebra ${\cal A}$ in third factor; realization of the 
Borel subalgebra ${\cal B}_-$ is sufficient.
Let us identify the generators $x_0,x_1\in{\cal B}_-$ of this Borel
subalgebra with the integrals of the exponential fields 
\eqn\xvert{
x_0={1\over q-q^{-1}}\int_0^{2\pi}V_-(u)\,  du\, ,
\qquad x_1={1\over q-q^{-1}}\int_0^{2\pi}V_+(u)\,  du\ .
}
One can check that these generators satisfy \FF\ the Serre relations
\serre. To do this one should express the fourth order products 
of $x$'s in \serre\ in terms of the ordered integrals of products of the
exponential fields $V_\pm(u)$\ . The calculations are simple but rather 
technical. We present them in the Appendix~A.

Substituting the expressions  \xvert\  for the 
generators $x_0,x_1$ into the ``reduced"
 universal $R$-matrix
$\overline{{\cal R}}$
\ \rseries , \rexpan \ 
one obtains a vector in 
${\cal B}_+$ whose coordinates are operators acting in 
the quantum space \mju.  It is natural to expect that it coincides with 
the \ ${\cal P}$-ordered exponent from  \Lopg .

\bigskip
\noindent
{\bf Conjecture.}\foot{This statement requires no
restrictions on the value of the central element $k$.}
{ \it The specialization of the 
``reduced" universal $R$-matrix ${\overline{\cal R}}\in
{\cal B}_+\otimes {\cal B}_-$\ \rseries\  for the case when 
$x_0, x_1\in {\cal B}_-$\  are realized as in  \xvert\ 
coincide with the ${\cal P}$-exponent 
\eqn\ex{\overline{\cal L}=
{\cal P}\exp\Big( \int_0^{2\pi} {\cal K}(u)\, du \Big)\ ,} 
where ${\cal K}(u)$ is defined in \calK.}
\bigskip

\noindent
One can check this conjecture in a few lowest orders in the 
series expansion for  the universal \RM. Substitute \xvert\ 
into \rexpan. It is not difficult to see that every polynomial of
$x$'s appearing in \rexpan\ as  a coefficient to the monomial in $y's$
can be written as a single  ordered
integral of the vertex operators (rather than their linear
combination). For example, the second order terms read 
\eqn\Xes{\eqalign{
\ds x_0^2={[2]_q\over q(q-q^{-1})^2}\ J(-,-)\, ,\qquad
&(x_0 x_1 -q^{-2} x_1 x_0)={[2]_q \over (q-q^{-1})}\ J(+,-)\, ,\cr
\ds x_1^2={[2]_q\over q(q-q^{-1})^2}\ J(+,+)\, ,\qquad
&(x_1 x_0 -q^{-2} x_0 x_1)={[2]_q  \over (q-q^{-1})}\ J(-,+)\, ,\cr}
}
where\foot{Note that this definition differs
by the factor $q^n$  from that given in Eq.(2.31) of Ref. \BLZZ.}
\eqn\Js{
J(\s_1,\s_2,\ldots,\s_n)=
\int_{2\pi\ge u_1\cdots u_{n}\ge0}
V_{\s_1}(u_1)\, V_{\s_2}(u_2)\cdots
V_{\s_n}(u_n)\ du_1\ldots du_{n}\ ,}
with $\s_i=\pm1$. Using \xvert\ and \Js\ one can rewrite the RHS of\
\rexpan\
as
\eqn\Lexpan{{
\overline{\cal R}}=1+y_0\ J(-)+y_1\ J(+) +y_0^2\ J(-,-) +y_1^2\
  J(+,+) + y_0 y_1\ J(-,+)
+ y_1 y_0 \ J(+,-)+ \ldots\ , }
which coincides with the first three terms of the expansion of 
\ ${\cal  P}$-exponent\ \ex.
We have verified this conjecture  
to within the terms of the fourth order
in the generators $x_0$ and $x_1$ (see Appendix~A).
Notice that starting from the fourth order one has to take into account
the Serre relations \serre.
The above conjecture suggests that   
the operators \ex\ can be reexpressed through algebraic
combinations of the two
elementary  integrals \xvert\ instead of the ordered
integrals \Js\foot{Perhaps this statement is less trivial than it might
appear. In fact, one can always write any product 
of $x_0$ and $x_1$
from \xvert\ as a linear combination of the integrals\ \Js, but
not vise versa, since the elementary integrals \xvert\ are algebraically
dependent due to the Serre relations.}.
Conversely, this statement combined with the uniqueness\ \KT\
of universal\ $R$-matrix
satisfying\ \imbed\ and \defineR\ implies the above conjecture. 

Finally let us stress that our proof of the \YBE s \yab\ and \ybeL\  is 
independent of this conjecture.

\newsec{ Commuting ${\bf T}$- and ${\bf Q}$-operators}

It is well known and simple consequence of the Yang-Baxter relation
\yab\ that appropriately defined traces of the operator matrices 
${\bf L}_j (\l)$ give rise to the operators ${\bf T}_j (\l)$ which 
commute for different values of the parameter $\l$, i.e.
\eqn\tjcom{
[{\bf T}_j(\l),{\bf T}_{j'}(\l)]=0\ .
}
In fact, there is an
certain ambiguity in the construction of these operators. Below we show
that this ambiguity is eliminated if we impose additional requirement
that the operators ${\bf T}_j (\l)$ also commute with the local IM
\loim,
\eqn\tjikcom{
[{\bf T}_j (\lambda), {\bf I}_{2k-1}]=0\ .
}
It is easy to check that 
the \YBE\ \yab\ is not affected if one multiplies the \LO\ \Lop\
by an exponent of the Cartan element  $H$
\eqn\Lf{
{\bf L}(\l) \lra {\bf L}^{(f)}(\l)=e^{ifH}{\bf L}(\l)\ ,
}
where $f$ is an arbitrary constant. Therefore the operators 
\eqn\Tf{
{\bf T}^{(f)}_j(\l)=\Tr_{\pi_j} \big[\ e^{ifH}{\bf L}_j(\l)\  \big]
}
satisfy the commutativity relations \tjcom\ for any value of $f$.
Moreover, this commutativity is not violated even if the
quantity $f$ is a function of $P$ rather than a constant (despite the
fact that in this case the operators \Lf\ do not necessarily satisfy 
the ordinary \YBE\ \yab). This is obvious if one uses the standard  
realization of the spin-$j$ representations $\pi_j$ of the algebra \uals\
\eqn\pij{
\pi_j[E]\,|k\rangle=[k]_q\,[2j-k+1]_q\, |k-1\rangle\, ,\quad
\pi_j[F]\,|k\rangle=|k+1\rangle\, ,\quad
\pi_j[H]\,|k\rangle=(2j-2k)\,|k\rangle\, ,
}
where $[k]_q=(q^k-q^{-k})/(q-q^{-1})$ and  the vectors 
$|k\rangle$\ $(k=0,1,\ldots,2j)$ 
form a basis  in the $(2j+1)$-dimensional space. 
Then, using \vcomm\ it is easy to show that 
all the diagonal entries of the $(2j+1)\times(2j+1)$ matrices 
${\bf L}_j(\l)$ commute with the operator $P$. As an immediate
consequence the quantity $f=f(P)$ in \Tf\ can be treated  
as a constant and therefore the commutativity \tjcom\ remains valid.
It follows also that the operators \Tf\ invariantly
act in each Fock module ${\cal F}_p $. 

The commutativity \tjikcom\ requires a special choice of the function
$f=f(P)$.  We show in the Appendix~C that the operators \Tf\
commute with\ $I_1 = L_0 -c/24$\ if 
\eqn\fp{
f=\pi\,  (P + N)\ .
}
Here $N$ is an arbitrary integer which obviously has no other effect 
on \Tf\ than the overall sign of this operator; in what follows we set
$N=0$ and define
\eqn\Tj{
{\bf T}_j(\l)=\Tr_{\pi_j}\big[\ e^{i\pi P H}{\bf L}_j(\l)\ \big]\  .
}
In fact, with this choice of $f$ the operators \Tj\ commute with
all the local IM \loim. This is demonstrated in Appendix~C. The
operators \Tj\ act invariantly in each Fock module ${\cal F}_p$
and satisfy both \tjcom\ and \tjikcom.  

The above operators ${\bf T}_j (\lambda)$ are CFT analogs of the 
commuting transfer-matrices of the Baxter's lattice theory. Besides
these commuting transfer-matrices the ``technology'' of the solvable
lattice models involves also another important object - the Baxter's
$Q$-matrix \Baxn. It turns out that another specialization of the
general ${\cal L}$ operator \Lopg\ leads to the CFT analog of the
$Q$-matrix \BLZZ.

Consider the so-called $q$-oscillator algebra generated by
the elements ${\cal H}\, ,{\cal E}_+ ,\, 
{\cal E}_- $\ subject to the relations 
\eqn\qosc{ q\, {\cal E}_{+}{\cal E}_{-} - q^{-1} {\cal E}_{-}{\cal E}_{+} = 
{1\over {q-q^{-1}}}\, , \qquad [{\cal H}, {\cal E}_{\pm}] = 
{\pm}\,2\,{\cal E}_{\pm}\ .}
One can easily show that the following
two maps of
the Borel subalgebra ${\cal B}_-$ of
$U_q\big(\widehat{sl}(2)\big)$ into the
q-oscillator  algebra \qosc\
\eqn\qmap{
h=h_0=-h_1\ra \pm {\cal H}\, ,\qquad y_0\ra \l{\cal E}_\pm\, ,\qquad
y_1\ra \l{\cal E}_\mp}
(here one has to choose all the upper or all the lower signs)
are {\it homomorphisms}.   
Under
these homomorphisms the operator \Lopg\ becomes an element of the 
algebra \qosc\
\eqn\lmap{
{\cal L}\ra {\bf L}_\pm(\l)=e^{\pm  i\pi P {\cal H}}\ {\cal P} 
\exp \Big( \lambda \int_{0}^{2\pi} du\, 
( V_{-}(u)\, q^{\pm\frac{{\cal H}}{ 2}} 
{\cal E}_{\pm} +
V_{+}(u)\, q^{\mp\frac{{\cal H}}{2}} {\cal E}_{\mp})\Big)\ .}
Let $\rho_{\pm}$ be any representations of \qosc\ such that the trace
\eqn\zpm{
Z_{\pm}(p)=\Tr_{\rho_{\pm}}[\ e^{\pm 2 \pi i p{\cal H}}\ ]}
exists  and does not vanish 
for complex $p$ belonging to the lower half plane, $\Im m\,
p<0$. Then define two operators
\eqn\Adef{
{\bf A}_\pm (\l)= Z^{-1}_{\pm}(P)\, \Tr_{\rho_{\pm}} [\ e^{\pm \pi i P{\cal
H}} {\bf L}_\pm(\l) \ ]\ .
}
Since we are interested in
the action of  these operators in ${\cal F}_p$ the operator 
$P$ in \Adef\ can be substituted by its eigenvalue $p$. The 
definition \Adef\ applies to the case
$\Im m\, p<0$.
However the
operators ${\bf A}_\pm (\l)$ can be defined for all complex $p$ (except
for some set of singular points on the real axis) by an analytic 
continuation in $p$. As was shown in \BLZZ, the trace in \Adef\ is
completely determined by the commutation relations \qosc\ and the
cyclic property of the trace, so the specific choice of the
representations $\rho_{\pm}$ is not significant as long as the above
property is maintained. 

The $Q$-operators 
(the CFT analogs of the Baxter's $Q$-matrix) are defined 
as 
\eqn\Qdef{
{\bf Q}_\pm(\l)=\l^{\pm2P/\beta^2}\, {\bf A}_\pm(\l)\ .
}
Similarly to the ${\bf T}$-operators they act invariantly in each Fock
module ${\cal F}_p$
\eqn\Qact{
{\bf Q}_\pm(\l): \qquad {\cal F}_p\rightarrow {\cal F}_p \ ,}
and commute with the local IM \loim. The operators ${\bf Q}_\pm(\l)$ 
with different  values of $\l$
commute among themselves and with all the operators ${\bf
T}_j(\l)$ 
\eqn\Qcom{[{\bf Q}_{\pm}(\lambda), {\bf Q}_{\pm}(\lambda')] = 
[{\bf Q}_{\pm}(\lambda),
{\bf T}_j (\lambda')] = 0\ .}
This follows from the appropriate specializations of the Yang-Baxter
equation \ybeL.

The operators ${\bf T}_j(\l)$ and ${\bf A}_\pm(\l)$ enjoy remarkable
analyticity properties as the functions of the variable $\l^2$. Namely,
all the matrix elements and eigenvalues of these operators are entire
functions of this variable\ \refs{\BLZ,\BLZZ}. The proof is carried out in
the Appendix~B. It is based on the result of \FSLS\ on the 
analyticity of certain Coulomb partition
functions which was obtained through
the Jack polynomial technique. Currently there is a complete proof of the
above analyticity for ${\bf T}_j(\l)$ for all values of $\beta^2$ in 
the domain $0 < \beta^2 < {1/2}$ (which corresponds to \cent \ ) and 
``almost complete'' proof  of this analyticity for  ${\bf
A}_\pm(\l)$ which extends to all rational values of $\beta^2$ and 
to almost all irrational values of $\beta^2$ (i.e. to all
irrationals values except for some set of the Lebesgue measure zero, see
Appendix~B for the details) in the above interval. It is natural to
assume that this analyticity remains valid for
those exceptional irrationals as well. 

\newsec{The functional relations}

It is well know from the lattice theory that  analyticity of the 
commuting transfer matrices become extremely powerful condition when 
combined with the functional relations which the transfer
matrices satisfy, and, 
in principle, allows one to determine all the eigenvalues.
Therefore, the functional
relations (FR) for the  operators ${\bf Q}_\pm(\l)$
and ${\bf T}_j(\l)$ are of primary interest. We start our consideration
with the ``fundamental'' FR (fundamental in the sense that it 
implies all the other functional 
relations involving the operators ${\bf T}_j(\l)$ or ${\bf Q}_\pm(\l)$).

{\sl i.) Fundamental Relation.}
 { The ``transfer-matrices'' ${\bf T}_j(\lambda)$ can be expressed in
terms of ${\bf Q}_{\pm}(\lambda)$ as \BLZZ\
\eqn\TQfund{2i\ \sin(  2\pi P  )
\ {\bf T}_j (\lambda) = {\bf Q}_{+}(q^{j+{1\over 2}}\lambda)
{\bf Q}_{-}(q^{-j-{1\over 2}}\lambda) -
{\bf Q}_{+}(q^{-j-{1\over 2}}
\lambda){\bf Q}_{-}(q^{j+{1\over 2}}\lambda)\ ,}
where $j$ takes  (half-) integer values
$j=0,{1\over2},1,{3\over2},\ldots,\infty$.}

Before going into the proof of \TQfund\ let us mention 
its simple but important corollary 

{\sl  ii.) $T$-$Q$ Relation.} 
The operators ${\bf Q}_{\pm}(\lambda)$ satisfy the Baxter's $T$-$Q$
equation
\eqn\TQ{{\bf T}(\lambda){\bf Q}_{\pm}(\lambda) = {\bf Q}_{\pm}(q\lambda)
+ {\bf Q}_{\pm}(q^{-1}\lambda)\ .}
where ${\bf T}(\l)\equiv{\bf T}_{1\over2}(\l)$. This equation  
can be thought of as the finite-difference analog of
a second order differential 
equation so we expect it to have two linearly independent
solutions. As ${\bf T}(\lambda)$ is a single-valued function of 
$\lambda^2$, i.e. it is a periodic 
function of $\log \lambda^2$, the operators
${\bf Q}_{\pm}(\lambda)$ are interpreted as two ``Bloch-wave'' 
solutions to the
equation \ \TQ. The operators
${\bf Q}_{\pm}(\lambda)$ satisfy the 
``quantum Wronskian'' condition
\eqn\wron{{\bf Q}_{+}(q^{1\over 2}\lambda)
{\bf Q}_{-}(q^{-{1\over 2}}\lambda) - 
{\bf Q}_{+}(q^{-{1\over 2}}\lambda){\bf Q}_{-}(q^{1\over 2}\lambda) = 
2i\  \sin ( 2\pi P )\ ,}
which is just a particular case of \TQfund\ with $j=0$.

To prove these relations consider   
more general ${\bf T}$-operators which correspond to the 
infinite dimensional highest weight representations of $U_q(sl(2))$. 
These new ${\bf T}$-operators are defined by the same formula as \Tj 
\eqn\Tjplus{
{\bf T}_j^+(\l)=\Tr_{\pi_j^+}\big[e^{i\pi P H}{\bf L}_j^+(\l)\big],\qquad
{\bf L}_j^+(\l)=\pi_j^+\big[ {\bf L}(\l)\big]} 
except that the trace is now  taken over the infinite dimensional
representation $\pi_j^+$ of \uals. The corresponding representation
matrices  
$\pi_j^+[E]$, $\pi_j^+[F]$ and
$\pi_j^+[H]$ for the generators of \uals\ are defined by the 
equations 
\eqn\spinj{ 
\pi_j^+[E]\,|k\rangle=[k]_q\,[2j-k+1]_q\,|k-1\rangle\,  ,\quad
\pi_j^+[F]\,|k\rangle=|k+1\rangle\, ,\quad
\pi_j^+[H]\,|k\rangle=(2j-2k)\,|k\rangle\, ,
}
which are similar to \pij, but the basis $|k\rangle$,\  is now infinite, 
$k=0,1,\ldots,\infty$.
The highest weight $2j$ of the representation $\pi_j^+$, 
$$
\pi_j^+(H)\,  |0\rangle =2j\, |0\rangle\, ,
$$
can take arbitrary complex values. Since we are interested in
the action of  the operators ${\bf T}_j^+(\l)$ in ${\cal F}_p$ the operator 
$P$ in \Tjplus\ can be substituted by its eigenvalue $p$. Similarly
to \Adef\ the 
definition \Tjplus\ makes sense only if $\Im m\, p<0$, but it
can be extended to all complex $p$ (except for some set of singular points on
the real axis) by the analytic continuation in $p$.
The operators \Tjplus\ thus defined satisfy the commutativity
conditions 
$$
[{\bf T}_j(\l),{\bf T}_{j'}^+(\mu)]=
[{\bf T}_j^+(\l),{\bf T}_{j'}^+(\mu)]=0\, ,
$$
which follow from the  appropriate specializations of the Yang-Baxter
equation \ybeL.

If $j$ takes  a non-negative integer or half-integer value the
matrices\ $\pi_j^+[E]$, $\pi_j^+[F]$ and
$\pi_j^+[H]$\ in \spinj\ have a
block-triangular form with two diagonal blocks, one (finite) being 
equivalent to the 
$(2j+1)$ dimensional representation $\pi_j$ and the
other (infinite) coinciding with the highest weight representation 
$\pi_{-j-1}^+$. Hence the following simple relation holds,
\eqn\simple{
{\bf T}_j^+(\l)={\bf T}_j(\l)+{\bf T}_{-j-1}^+(\l)\, ,
\qquad j=0,1/2,1,3/2,\ldots\ .
}
In some ways the operators ${\bf T}_j^+(\l)$ are simpler 
than ${\bf T}_j(\l)$. Making a similarity transformation 
$$ 
E \ra \l\, E\, , \qquad F\ra \l^{-1} F\, ,
$$
which does not affect the trace in \Tjplus\ and observing
that 
$$
 \lambda^2\, \pi^+_j(E)\, |k\rangle={[k]_q\over q-q^{-1}}\ 
\big(\lambda_+^2\,  q^{-k}-
\lambda_-^2\,  q^{k}\, \big)\, |k-1\rangle\, ,
\qquad k=0,1,\ldots,\infty\, ,
$$
where 
\eqn\lpm{
\l_+=\l\,  q^{j+\hf}\, ,\qquad \l_-=\l\,  q^{-j-\hf}\, ,
} 
one can conclude that the operator ${\bf T}_j^+(\l)$ can be written as
\eqn\depen{
{\bf T}_j^+(\l)= {\bf T}_j^+(0)\ \Phi(\l_+,\l_-)\, ,
}
where 
\eqn\norma{
{\bf T}_j^+(0)={e^{2\pi i (2j+1)  P}\over 2i \sin (2\pi P)}
}
and $\Phi(\l_+,\l_-)$ is a series in $\l_+^2$ and $\l_-^2$ with the
coefficients which do not depend on $j$ and the the leading coefficient being
equal to 1. Remarkably, the expression \depen\ further 
simplifies since the quantity  $\Phi(\l_+,\l_-)$ factorizes into a product
of two operators \Adef
\eqn\factor{
2i \sin(2\pi P)\ {\bf T}_j^+(\l)=e^{2\pi i (2j+1)  P}\ 
{\bf A}_+\big(\l\, q^{j+\hf}\big)\,{\bf A}_-\big( \l\,  q^{-j-\hf}\big)\  .
}
This factorization can be proved algebraically 
by using decomposition properties of the
tensor product of
two representations of the $q$-oscillator algebra (the latter are also
representations of the Borel subalgebra of 
$U_q\big(\widehat{sl}(2)\big)$ with
respect to the co-multiplication from 
$U_q\big(\widehat{sl}(2)\big)$. The detail of the calculations are 
presented in the Appendix~D. The functional relation \ \TQfund\ 
trivially follows from \simple\  and 
\ \factor.

The relation \wron\ shows that the operators ${\bf Q}_{+}(\l)$ and 
${\bf Q}_{-}(\l)$ are functionally dependent.
Using this dependence one can write \TQfund\  as 
\eqn\TjQb{
{\bf T}_j(\l)={\bf Q}(q^{j+\hf}\,\l)\,{\bf Q}(q^{-j-\hf}\,\l)
\sum_{k=-j}^j {1\over {\bf Q}(q^{k+\hf}\,\l)\,{\bf Q}(q^{k-\hf}\,\l)\, ,}
}
where ${\bf Q}(\l)$ is any one of ${\bf Q}_+(\l)$ and  ${\bf Q}_-(\l)$.

The last group of FR we want to discuss here is the relations
involving solely the transfer matrices ${\bf T}_j(\l)$ and usually referred
to as the ``fusion relations''\foot{In fact, all the above FR can also
be called the fusion relation since they all follow from
\factor\ which describes the ``fusion'' of the q-oscillator algebra
representations.} \KR. Note that these are again simple
corollaries of the ``fundamental relation'' \TQfund.

{\sl iii.) Fusion relations.} The transfer matrices  ${\bf T}_j(\l)$
satisfy 
\eqn\funcrel{
{\bf T}_j(q^\hf\,\l)\, {\bf T}_j(q^{-\hf}\,\l)
=1+{\bf T}_{j+\hf}(\l)\,{\bf T}_{j-\hf}(\l)\, ,
}
where ${\bf T}_0(\l)\equiv 1$. 
These can also be equivalently rewritten 
as 
\eqn\fusa{
{\bf T}(\l) \, {\bf T}_j(q^{j+\hf}\,\l)={\bf T}_{j-\hf}(q^{j+1}\,\l)+
{\bf T}_{j+\hf}(q^j\,\l)\, , 
}
or as 
\eqn\fusb{
{\bf T}(\l) \, {\bf T}_j(q^{-j-\hf}\,\l)={\bf T}_{j-\hf}(q^{-j-1}\,\l)+
{\bf T}_{j+\hf}(q^{-j}\,\l)\ .
}

Considerable reductions of the FR occur when $q$ is a root of unity. Let
\eqn\qroot{
q^N=\pm1 \qquad {\rm and } \qquad q^n\not=\pm 1 \qquad {\rm for\ any\ 
integer} \qquad 0<n<N,
}
where  $N\ge2$ is some integer. When using \TQfund\ it is easy to
obtain that  
\eqn\tsum{
e^{2\pi i NP}\,  {\bf T}_j(\l)+{\bf T}_{{N\over2}-1-j}(\l
q^{N\over2})={\sin(2\pi NP)\over\sin(2\pi P)}\
{\bf Q}_+(\l q^{j+{1\over2}}) 
{\bf Q}_-(\l q^{-j-{1\over2}}) }
for $j=0,{1\over2},\dots,{N\over2}-1$.
Similarly,
\eqn\talone{
{\bf T}_{{N\over2}-{1\over2}}(\lambda)={\sin(2\pi NP)\over\sin(2\pi P)}\,
{\bf Q}_+(\l q^{N\over2}){\bf Q}_-(\l q^{N\over2})\ .
}
Moreover in this case 
there is an extra relation involving only ${\bf T}$-operators
\eqn\TN{
{\bf T}_{{N\over2}}(\l)=2\cos(2\pi NP)+{\bf T}_{{N\over2}-1}(\l)\ ,}
as it readily follows from \TQfund.
As is shown in \BLZZZ\ this allows to bring the FR \funcrel\ to the form
identical to the functional TBA equations  (the $Y$-system) of the 
$D_N$ type. 

Additional simplifications occur when the operators ${\bf T}_j$ act
in Fock spaces ${\cal F}_p$ with special values of $p$,
\eqn\pmn{p={\ell+1\over {2\, N}}\ ,}
where $\ell\geq 0$ 
is an integer such that $2p\not= n\beta^2 +m$ for any 
integers $n$ and $m$. Then the RHS's of \tsum\ and \talone\ vanish and these
relations lead to
\eqn\Tredm{\eqalign{{\bf T}_{{N\over2}-j-1}(\lambda) 
= (-1)^{\ell}&\  {\bf T}_j (q^{N\over2}\,\lambda)\, , \quad
\hbox{for} \quad \textstyle{j=0, \half , 1, ..., {N\over2}-1}\, ;\cr
&{\bf T}_{{N\over2}-{1\over2} }(\lambda) =
0\ .}}
Further discussion of this case can be found in \BLZ\ and \BLZZZ.

Finally, some remarks concerning the lattice theory are worth making.
Although our construction of the $Q$-operators in terms of the
$q$-oscillator representations was given here specifically for the case of
continuous theory, it is clear that the lattice $Q$-matrices admit
similar construction. In particular the $Q$-matrix 
of the six-vertex model 
can be obtained as a transfer matrix associated with 
infinite dimensional
representations of the $q$-oscillator algebra \qosc
\foot{Using this construction it is possible, in particular, to
reproduce a remarkably simple expression for an 
arbitrary matrix element of the $Q$-matrix of the zero field six-vertex
model in the ``half-filling'' sector\ \Baxn. }. In the case of the 
six-vertex vertex model with nonzero (horizontal) field this
construction gives rise to two $Q$-matrices, ${\bf Q_{\pm}}$.
As the 
structure of the FR \TQfund,\ \TQ,\ \factor\ 
is completely determined
by the decomposition properties of products of representations of
$U_q\big(\widehat{sl}(2)\big)$, all these FR are valid in the lattice
case, with minor modifications mostly related to the normalization 
conventions 
of the lattice transfer matrices.

\newsec{Acknowledgments} The authors thanks R.~Askey and 
D.S.~Libinsky for bringing papers \refs{\HL,\driver} to our attention.
V.B. thanks L.D.Faddeev, E. Frenkel and S.M.Khoroshkin 
for interesting discussions and 
the Department of Physics and Astronomy, Rutgers University 
for hospitality.

\appendix{A}{}

\def\ds{\displaystyle}
\def\YBE{Yang-Baxter equation}
\def\LO{${\bf L}$-operator}
\def\RM{$R$-matrix}

\def\veps{\varepsilon}

\def\d{\delta}
\def\a{\alpha}
\def\b{\beta}
\def\l{\lambda}

\def\s{\sigma}

\def\lra{\longrightarrow}
\def\ra{\rightarrow}

\def\hf{{1\over2}}
 
\def\ds{\displaystyle}
\def\Tr{{\rm Tr}}

\def\crx{\cr\noalign{\medskip}}
\def\xtra{\noalign{\medskip}}
\def\smatrix#1{\left[\matrix{#1}\right]}
\def\qa{(q-q^{-1})}
\def\cite#1{#1}

Here we present the results on the series expansion verification of our
conjecture on the structure of the universal $R$-matrix for the quantum 
Kac-Moody algebra $U_q(\widehat{sl}(2))$.

We will need expressions for  products of the basic contour integrals 
\xvert\ in terms of linear combinations of the ordered integrals \Js.
To derive them one only has  to use the commutation relation 
\vcomm\ for the vertex operators. For example, consider the 
second order product
\eqn\xzo{\eqalign{
(q-q^{-1})^2\  x_0 x_1 =&\int_0^{2\pi}\int_0^{2\pi} :e^{-2\varphi(u_1)}:\ 
:e^{2\varphi(u_2)}: du_1 du_2\cr
\noalign{\vskip 0.3 cm}
=&\int_{2\pi>u_1>u_2>0}\ldots\, du_1du_2+
\int_{2\pi>u_2>u_1>0}\ldots\, du_1du_2\cr
\noalign{\vskip 0.3 cm}
=&J(-,+)+q^2 J(+,-)\ ,\cr}}
where $J$'s are defined in \Js.

For the $n$-th order products one has to split the domain of integration in 
$n$-tuple integral into $n!$ pieces corresponding to all possible
orderings of the integration variables and then rearrange the  products
of the vertex operators using the commutation relations \vcomm. Below
we present the results of these calculations for the products of
orders less or equal to four, 
\eqn\xpow{\eqalign{
x_0&={1\over \qa}\,J(-)\, ,\cr
x_0^2&={q^{-1}[2]_q\over \qa^2}\ J(-,-)\, ,\cr
x_0^3&={q^{-3}[2]_q [3]_q\over \qa^3}\ J(-,-,-)\, ,\cr
x_0^4&={q^{-6}[2]_q[3]_q[4]_q\over \qa^4}\ J(-,-,-,-)\, ,}\qquad
\eqalign{x_1&={1\over \qa}\,J(+)\, ,\cr
x_1^2&={q^{-1}[2]_q\over \qa^2}\ J(+,+)\, ,\cr
x_1^3&={q^{-3}[2]_q [3]_q\over \qa^3}\ J(+,+,+)\, ,\cr
x_1^4&={q^{-6}[2]_q[3]_q[4]_q\over \qa^4}\ J(+,+,+,+)\, , }
}
\eqn\xzo{
\smatrix{x_0x_1\crx x_1x_0\cr}={1\over \qa^2}\smatrix{1&q^{2}\crx 
q^{2}&1\cr}
\smatrix{J(-,+)\crx J(+,-)\cr}\, ,}
\eqn\xzzo{
\smatrix{x_0^2x_1\crx x_0x_1x_0\crx x_1x_0^2}=
{[2]_q\over q\qa^3}\smatrix{1&q^2&q^4\crx q^2&q^2&q^2\crx q^4&q^2&1\cr}
\smatrix{J(-,-,+)\crx J(-,+,-) \crx J(+,-,-)\cr}\, ,}
\eqn\xooz{
\smatrix{x_1^2x_0\crx x_1x_0x_1\crx x_0x_1^2}=
{[2]_q\over q\qa^3}\smatrix{1&q^2&q^4\crx q^2&q^2&q^2\crx q^4&q^2&1\cr}
\smatrix{J(+,+,-)\crx J(+,-,+) \crx J(-,+,+)\cr}\, ,}
\eqn\xzzzo{
\left[\matrix{x_0^3 x_1\cr\xtra x_0^2x_1 x_0\cr \xtra
x_0 x_1 x_0^2\cr\xtra x_1x_0^3}\right]= 
{[2]_q\over \qa^4}
\ \left[\matrix{q^{-3}[3]_q&q^{-1}[3]_q&q\,[3]_q&q^3[3]_q\cr\xtra
         q^{-1}[3]_q& q+2q^{-1}&2q+q^{-1}&q\,[3]_q\cr\xtra
         q\,[3]_q&2q+q^{-1}&q+2q^{-1}&q^{-1}[3]_q\cr\xtra
         q^3[3]_q&q\,[3]_q&q^{-1}[3]_q&q^{-3}[3]_q\cr}\right]
\left[\matrix{J(-,-,-,+)\cr\xtra J(-,-,+,-)\cr\xtra
J(-,+,-,-)\cr\xtra J(+,-,-,-)\cr}\right]\, ,
}

\eqn\xoooz{
\left[\matrix{x_1^3 x_0\cr\xtra x_1^2x_0 x_1\cr \xtra
x_1 x_0 x_1^2\cr\xtra x_0x_1^3}\right]= 
{[2]_q\over \qa^4}
\ \left[\matrix{q^{-3}[3]_q&q^{-1}[3]_q&q\,[3]_q&q^3[3]_q\cr\xtra
         q^{-1}[3]_q& q+2q^{-1}&2q+q^{-1}&q\,[3]_q\cr\xtra
         q\,[3]_q&2q+q^{-1}&q+2q^{-1}&q^{-1}[3]_q\cr\xtra
         q^3[3]_q&q\,[3]_q&q^{-1}[3]_q&q^{-3}[3]_q\cr}\right]
\left[\matrix{J(+,+,+,-)\cr\xtra J(+,+,-,+)\cr\xtra
J(+,-,+,+)\cr\xtra J(-,+,+,+)\cr}\right]\, ,
}
\eqn\xzozo{
\smatrix{x_0^2x_1^2\crx x_0 x_1 x_0 x_1\crx
         x_0x_1^2x_0\crx x_1x_0^2x_1\crx x_1x_0x_1x_0\crx x_1^2x_0^2\cr}=
{q^2[2]_q^2\over \qa^4}
\smatrix{q^{-4}&q^{-2}&1&1&q^2&q^4\crx
         q^{-2}&{2q^{-1}\over [2]_q}&1&1&{2q\over [2]_q}&q^2\crx
         1&1&[3]_q-2&1&1&1\crx
         1&1&1&[3]_q-2&1&1\crx
         q^2&{2q\over[2]_q}&1&1&{2q^{-1}\over[2]_q}&q^{-2}\crx
         q^4&q^2&1&1&q^{-2}&q^{-4}\crx}                       
\smatrix{J_{12}\crx J_{13}\crx J_{14}\crx
         J_{23}\crx J_{24}\crx J_{34}\cr}\ ,
}
where
\eqn\JJJ{\eqalign{
&J_{12}=J(-,-,+,+)\, ,\qquad J_{13}=J(-,+,-,+)\, 
,\qquad J_{14}=J(-,+,+,-)\, ,\cr
&J_{23}=J(+,-,-,+)\, ,\qquad J_{24}=J(+,-,+,-)\, ,
\qquad J_{34}=J(+,+,-,- )\, .\cr}}
We can now invert  most of these relations (except \xzzzo\ and \xoooz) to
express $J$'s in terms of products of $x$'s. This is not possible for
\xzzzo\ and \xoooz\ because the products of $x$'s in the left hand
sides are linearly dependent (the rank of the four by four
matrix therein is equal to three) as a manifestation of the Serre relations.
In fact, using \xzzzo\ and \xoooz\ one can easily check that 
the Serre relations
\serre\ for the basic contour integrals \xvert\ are indeed
satisfied. It is, perhaps,  not surprising that the $J$-integrals
entering \xzzzo\ and \xoooz\ appear in the expansion of the\ \LO\
\Lopg\ only in certain linear combinations\foot{This happens again due
to the Serre relation but now for the generators $y_0$ and $y_1$.}
which can be expressed through
the products of $x$'s. We will need the following combinations,
\eqn\Jser{\eqalign{
&J(-,+,+,+)\ +\ J(+,+,+,-)={\qa^2 \over [2]_q^2}\bigg\{{3\over
[3]_q}x_0x_1^3-2x_1x_0x_1^2+x_1^2x_0x_1\bigg\}\, ,\crx
&J(+,-,+,+)-[3]_qJ(+,+,+,-)=\crx
&\phantom{J(-,+,-,-)[3]xxxx}={\qa^2 \over [2]_q^2}\bigg\{
-2x_0x_1^3+([3]_q+q^{-2})x_1x_0x_1^2-q^{-1}[2]_qx_1^2x_0x_1\bigg\}\, ,\crx
&J(+,+,-,+)+[3]_qJ(+,+,+,-)={\qa^2 \over [2]_q^2}\bigg\{
x_0x_1^3-q^{-1}[2]_qx_1x_0x_1^2+q^{-2}x_1^2x_0x_1\bigg\}\, ,\crx
&J(-,-,-,+)\ +\ J(+,-,-,-)={\qa^2 \over [2]_q ^2}\bigg\{{3\over
[3]_q}x_0^3x_1-2x_0^2x_1x_0+x_0x_1x_0^2\bigg\}\, ,\crx
&J(-,-,+,-)-[3]_qJ(+,-,-,-)=\crx
&\phantom{J(-,+,-,-)[3]xxxx}={\qa^2 \over [2]_q^2}\bigg\{
-2x_0^3x_1+([3]_q+q^{-2})x_0^2x_1x_0-q^{-1}[2]_qx_0x_1x_0^2\bigg\}\, ,\crx
&J(-,+,-,-)+[3]_qJ(+,-,-,-)={\qa^2 \over [2]_q^2}\bigg\{
x_0^3x_1-q^{-1}[2]_qx_0^2x_1x_0+q^{-2}x_0x_1x_0^2\bigg\}\, .\crx
}}
For the rest of $J$'s one has
\eqn\xoa{
\smatrix{J(-,+)\crx J(+,-)\cr}=
{ \qa\over [2]_q}\smatrix{-q^{-2}&1\crx1&- q^{-2}\cr}
\smatrix{x_0x_1\crx x_1x_0\cr}\, ,}

\eqn\xzo{
\smatrix{J(-,-,+)\crx J(-,+,-) \crx J(+,-,-)}=
{(q-q^{-1})\over[2]^2_q}
 \smatrix{q^{-2}&-q^{-1}[2]_q&1\crx -q^{-1}[2]_q&q^{-1}
[4]_q&-q^{-1}[2]_q\crx 1&-q^{-1}[2]_q&q^{-2}\cr}
\smatrix{x_0^2x_1\crx x_0x_1x_0\crx x_1x_0^2}\, ,
}

\eqn\xo{
\smatrix{J(+,+,-)\crx J(+,-,+) \crx J(-,+,+)}=
{(q-q^{-1})\over[2]^2_q}
 \smatrix{q^{-2}&-q^{-1}[2]_q&1\crx -q^{-1}[2]_q&q^{-1}
[4]_q&-q^{-1}[2]_q\crx 1&-q^{-1}[2]_q&q^{-2}\cr}
\smatrix{x_1^2x_0\crx x_1x_0x_1\crx x_0x_1^2}\, ,}

\eqn\iozo{
\eqalign{
J(+,+,-,-)&=q^{-2}{\qa\over [4]_q [2]_q}\bigg\{
q^{2}{2\over [2]_q}
x^2_0x^2_1-q^2[2]_q x_0x_1x_0x_1+ \qa x_0x^2_1x_0+
\cr&  \qa x_1x^2_0x_1+q^{-2}[2]_q x_1x_0x_1x_0-q^{-2}{2\over [2]_q}
x^2_1x^2_0\bigg\}\, ,
\cr
J(+,-,+,-)&=q^{-2}{\qa\over [4]_q [2]_q}\bigg\{
-q^2[2]_qx^2_0x^2_1+(2q+q^3+q^5)x_0x_1x_0x_1-\cr
&  \qa
[3]_q x_0x^2_1x_0-
\qa[3]_q x_1x^2_0x_1-
\cr&\ \ \ \ (q^{-5}+q^{-3}+2q^{-1}) x_1x_0x_1x_0+q^{-2}[2]_q
x^2_1x^2_0\bigg\}\, ,
\cr
J(+,-,-,+)&=q^{-2}{\qa^2\over [4]_q [2]_q}\bigg\{
x^2_0x^2_1-[3]_q x_0x_1x_0x_1+  x_0x^2_1x_0+
\cr& [3]_q x_1x^2_0x_1-[3]_q x_1x_0x_1x_0
+x^2_1x^2_0\bigg\}\, ,
\cr
J(-,+,+,-)&=q^{-2}{\qa^2\over [4]_q [2]_q}\bigg\{
x^2_0x^2_1-[3]_q x_0x_1x_0x_1+[3]_q x_0x^2_1x_0+
\cr&  x_1x^2_0x_1-[3]_q x_1x_0x_1x_0
+x^2_1x^2_0\bigg\}\, ,
\cr
J(-,+,-,+)&=q^{-2}{\qa\over [4]_q [2]_q}\bigg\{
q^{-2}[2]_qx^2_0x^2_1-(q^{-5}+q^{-3}+2q^{-1})x_0x_1x_0x_1-\cr
&  \qa
[3]_q x_0x^2_1x_0-
 \qa[3]_q x_1x^2_0x_1+
\cr&\ \ \ \ (2q+q^3+q^5) x_1x_0x_1x_0-q^2[2]_q
x^2_1x^2_0\bigg\}\, ,
\cr
J(-,-,+,+)&=q^{-2}{\qa\over [4]_q [2]_q}\bigg\{
-q^{-2}{2\over [2]_q}
x^2_0x^2_1+q^{-2}[2]_q x_0x_1x_0x_1+  \qa x_0x^2_1x_0+
\cr&  \qa x_1x^2_0x_1-q^2[2]_q  x_1x_0x_1x_0+q^2{2\over [2]_q}
x^2_1x^2_0\bigg\}\, .
}}
Expanding the ${\cal P}$-exponent in \ex\ in a series one obtains
\eqn\Lexp{
\exp\bigg( \int_0^{2\pi} {\cal K}(u) du \bigg) = 1+\sum_{n=1}^\infty 
\sum_{\{\sigma_i=\pm 1\}}  y_{\s_1}y_{\s_2}\cdots y_{\s_n}
\ J(-\s_1,-\s_2,\ldots,-\s_n)\, ,}
where 
$$
y_+=y_0\, ,\qquad y_-=y_1\ .
$$
Let us restrict our attention to the terms in \Lexp\ of the order four
or lower. One can exclude the products $y_0y_1^3$ and
$y_0^3y_1$ using the Serre relations \serre.
Then one can substitute the $J$-integrals 
in \Lexp\ with the corresponding expressions\ \iozo-\Jser.
There is no need 
to rewrite \Lexp\ again since this substitution  is rather mechanical 
and no cancelation can occur. The resulting expression is to be 
compared with the corresponding expansion of the universal $R$-matrix.
The latter can be found using the results of\  \refs{\KT- \KST}. The
notation for the generators of the  $U_q(\widehat{sl}(2))$ used in
that papers is  different from ours. The
generators $e_{\a},e_{-\a},e_{\b},e_{-\b}, h_{\a}, h_{\b}$ 
in \refs{\KT- \KST}\ 
are related to $x_0, x_1, y_0, y_1, h_0, h_1$ in \KMsimple-\serre\ as follows
\eqn\grel{\eqalign{
e_{\a}=q^{-h_0}y_0,\qquad &e_{-\a}=x_0q^{h_0},\qquad h_{\a}=h_0\, ,\cr
e_{\b}=q^{-h_1}y_1,\qquad &e_{-\b}=x_1q^{h_1},\qquad h_{\b}=h_1\, .\cr}}
The expression for the "reduced"  
universal \RM\ \rseries\ follows from Eq.(5.1) of Ref.\ \KST
\eqn\KTR{\eqalign{
{\overline {\cal R}}=&
\left(\prod_{n\geq 0}^{\rightarrow} \exp_{q^{-2}}
\left( (q-q^{-1})\ e_{\alpha+n\delta}\, q^{h_{\alpha+n\delta}}
 \otimes q^{-h_{\alpha+n\delta}}\,  e_{-\alpha-n\delta}
\right)\right)\times\crx &
\exp\left( \sum_{n>0}(q-q^{-1})\ {n (e_{n\delta}\, q^{h_{n\delta}}
\otimes q^{-h_{n\delta}}\,  e_{-n\delta})\over[2n]_q}\ \right) \times\crx &
\left(\prod_{n\geq 0}^{\leftarrow}\exp_{q^{-2}}\left(\  (q-q^{-1})\ 
e_{\beta+n\delta}\,q^{h_{\beta+n\delta}}  \otimes
q^{-h_{\beta}+n\delta}\, 
e_{-\beta-n\delta}\right)\right)\ ,\crx}}
where 
$$\exp_p(x)=\sum_{n=0}^\infty {p^{(n-1)(2-n)/2}\ x^n\over [n]_p!}\, ,
\qquad
[n]_p!=[1]_p[2]_p\cdots [n]_p\  $$
and $h_{\gamma+n\delta}=h_{\gamma}+n\, (h_{\alpha}+h_{\beta})\ (
h_{\gamma}=0,\ h_{\alpha},\ h_{\beta})$.
The index $n$ of the multipliers  increases
 from left to right in the first ordered product above and decreases 
in the second one. The root vectors $e_{\alpha+n\delta}$,
$e_{-\alpha-n\delta}$, etc. appearing  in \KTR\ are defined
recursively  by Eqs.(3.2)--(3.5) of Ref.\KST.
Applying these
formulae one obtains first few of them 
\def\ea{e_{\a}}
\def\eb{e_{\b}}
\def\fa{e_{-\a}}
\def\fb{e_{-\b}}
\eqn\rootva{\eqalign{
e_{\a+\d}&={1\over[2]_q}(\ea^2 \eb -(1+q^{-2}) \ea \eb \ea+q^{-2} \eb
\ea^2)\, , \crx 
e_{\b+\d}&={1\over[2]_q}(\ea\eb^2 -(1+q^{-2}) \eb \ea\eb+q^{-2} \eb^2
\ea)\, ,\cr }}
\eqn\rootvb{\eqalign{
e_{-\a-\d}&={1\over[2]_q}(\fb\fa^2 -(1+q^2) \fa \fb \fa+q^{2} \fa^2
\fb)\, ,\crx 
e_{-\b-\d}&={1\over[2]_q}(\fb^2\fa -(1+q^{2}) \fb \fa \fb+q^{2} \fa
\fb^2)\, ,\cr }}
\eqn\rootvc{
e_{\d}=\ea \eb -q^{-2} \eb \ea, \qquad
e_{-\d}=\fb \fa -q^{2} \fa \fb\, ,}
\eqn\rootve{\eqalign{
&e_{2\d}={1\over 2q^2 [2]_q}\bigg\{2q^2\ea^2\eb^2-q^2[2]_q^2\,\ea\eb\ea\eb
+(q^2-q^{-2})(\ea\eb^2\ea+\eb\ea^2\eb)+\crx
&\ \ \ \ \ \ \ \ \ \ \ \ \ \ q^{-2}[2]^2_q\,\eb\ea\eb\ea-
2q^{-2}\eb^2\ea^2\bigg\}\, ,\cr
&e_{-2\d}=-{q^2\over 2[2]_q}\bigg\{2q^2\fa^2\fb^2-q^2[2]_q^2\,\fa\fb\fa\fb+
\crx&(q^2-q^{-2})(\fa\fb^2\fa+\fb\fa^2\fb)+q^{-2}[2]^2_q\,\fb\fa\fb\fa-
2q^{-2}\fb^2\fa^2\bigg\}\, .\cr
}}
These formulae enable us to calculate the expansion of the universal 
\RM\ \KTR\ to within the forth order terms. Substituting
\rootva-\rootve\ into \KTR, expanding the the exponents
 and calculating their product
one gets precisely the result obtained above from the expansion 
of the \LO\ \Lopg\ given by \Lexp\ and\ \iozo-\Jser.

Finally notice the that the negative root vectors \rootvb-\rootve\ have
particular simple expressions in terms of $J$-integrals \Js, namely
they all reduce to just a single $J$-integral as one easily obtains
from \rootvb-\rootve\ and \iozo-\xo,
\eqn\uyt{\eqalign{
&q^{-h_{\delta}} e_{-\delta}=-q^4{[2]_q\over q-q^{-1}}\ J(+,-)\ ,\cr
&q^{-h_{\alpha+\delta}} e_{-\alpha-\delta}=q^6{[2]_q\over q-q^{-1}}
\ J(+,-,-)\ ,\cr
&q^{-h_{\beta+\delta}} e_{-\beta-\delta}=q^6{[2]_q\over q-q^{-1}}\
 J(+,+,-)\  ,\cr
&q^{-h_{2\delta}} e_{-2\delta}=-q^4{[2]_q[4]_q\over 2(q-q^{-1})}\
 J(+,+,-,-)\  .
}}
\appendix{B}{}

In this Appendix we show that for $0<\beta^2<1/2$ the 
operators ${\bf T}_j(\lambda)$\ \Tj\ and
${\bf A}_\pm(\lambda)$\ \Adef\ are entire functions of the variable
$\lambda^2$. 

Consider the simplest
 nontrivial ${\bf T}$-operator 
${\bf T}(\l)={\bf T}_{\hf}(\l)$ 
which corresponds to the two-dimensional representation of
$U_q(sl(2))$. In this case 
\eqn\spinhalf{
\pi_{1\over2}(E)=\pmatrix{0&1\cr0&0},\qquad
\pi_{1\over2}(F)=\pmatrix{0&0\cr1&0},\qquad
\pi_{1\over2}(H)=\pmatrix{1&0\cr0&-1}\ .
}
Using these expressions to compute  the trace in \Tj\ 
one finds
\eqn\Thf{
{\bf T}(\lambda)=2\cos (2\pi P) +
 \sum_{n=1}^{\infty}\lambda^{2n}\ {\bf G}_n\ ,
}
where
\eqn\Gn{
{\bf G}_n= q^n e^{2 i\pi P} J(\underbrace{-,+,\ldots,-,+}_{2n\;  {\rm
elements}})+ q^n e^{-2 i\pi P} J(\underbrace{+,-,\ldots,+,-}_{2n\;
{\rm elements}})} 
with $J$'s  defined in \Js. 
The operators ${\bf G_n}$ are the ``nonlocal integrals of motion''
(NIM) \BLZ\ which commute among themselves and with all operators 
${\bf T}_j(\l)$. They act invariantly 
in each Fock module ${\cal F}_p$. In particular, the vacuum state
$ |p\rangle\in {\cal F}_p$ is an eigenstate of all operators ${\bf G}_n$
\eqn\gvac{
{\bf G}_{n}\, |p\rangle=G_{n}^{(vac)}(p)\, |p\rangle\ ,}
where the eigenvalues $G_{n}^{(vac)}(p)$ are given by the integrals \BLZ\
\eqn\mcnaa{\eqalign{
G_{n}^{(vac)}(p)&=\int_{0}^{2\pi}du_1 \int_{0}^{u_1}dv_1
\int_{0}^{v_1}du_2 \int_{0}^{u_2}dv_2 ...
\int_{0}^{v_{n-1}}du_{n}\int_{0}^{u_n}dv_n\cr
\prod_{j>i}^{n}&{\Bigl(4\ \sin\big({{u_i - u_j}\over 2}\big)\
 \sin\big({{v_i - v_j}\over 2}\big)\Bigl)}^{2\beta^2}\
\prod_{j\geq i}^{n}{\Bigl(2\ \sin\big({{u_i - v_j}\over 2}\big)
\Bigl)}^{-2\beta^2}\cr
\prod_{j>i}^{n}&{\Bigl(2\ \sin\big(
{{v_i - u_j}\over 2}\big)\Bigl)}^{-2\beta^2}
\ \ 2\ \cos\Bigl(2 p\ \big(\pi+\sum_{i=1}^{n}
(v_i - u_i ) \big)\Bigl)\ .}}

Let us now examine the convergence properties of the series
\eqn\Tvac{
T^{(vac)}(\l)=\cos(2\pi p)+\sum_{n=1}^\infty \lambda^{2n}\
G_{n}^{(vac)}(p)}
for the vacuum eigenvalue of the operator ${\bf T}(\l)$. A similar
problem was studied in \FSLS\ for the series
\eqn\Zsal{
{\cal Z}(\l)=1 +\sum_{n=1}^\infty \lambda^{2n}\ {\cal Z}_n}
with
\eqn\Zn{\eqalign{
{\cal Z}_n&={1\over (n!)^2}\int_{0}^{2\pi}du_1
\int_{0}^{2\pi}du_2\cdots\int_0^{2\pi}du_n
\int_{0}^{2\pi}dv_1
\int_{0}^{2\pi}dv_2\cdots\int_0^{2\pi}dv_n\cr
\prod_{j>i}^{n}&{\Bigl|4\ \sin\big({{u_i - u_j}\over 2}\big)\
\sin\big({{v_i - v_j}\over 2}\big)\Bigl|}^{2\beta^2}\ 
\prod_{j,i=1}^{n}{\Bigl|2\ \sin\big({{u_i - v_j}\over 2}\big)
\Bigr|}^{-2\beta^2}\ ,\cr}}
where $0<\beta^2<1/2$.
It was shown (using the Jack polynomial technique) that the
leading asymptotics of the integrals \Zn\ for large $n$ is given by
\eqn\Zbound
{\log{\cal Z}_n=2\,  (\beta^2-1)\ n\,\log n +O(n)\, ,\ \ \ \ \
\ n\to\infty}
and hence 
series \Zsal\ defines an entire function of the variable $\lambda^2$. 
It is easy to see that 
\eqn\Gbound{
\big|G_{n}^{(vac)}(p)\big| <    {\cal Z}_n}
and therefore the eigenvalue \Tvac\ is also an entire function of
$\lambda^2$. Similar considerations apply to an arbitrary matrix
elements of ${\bf T}(\l)$ between the states in ${\cal F}_p$.
Thus all matrix elements and eigenvalues of ${\bf T}(\l)$
are entire functions\foot{
The higher spin operators  ${\bf T}_j(\lambda)$ with $j>1/2$ can
be polynomially expressed through ${\bf T}_{{1\over 2}}(\l)$ (as
it follow from
\fusa) and obviously enjoy the same analyticity properties.} 
 of $\l^2$.

Consider now the vacuum eigenvalue $A^{(vac)}(\l)$ 
of the operator ${\bf A}(\l)\equiv
{\bf A}_+(\l)$ defined in \Adef. It can be written as a series
\eqn\Avac{
A^{(vac)}(\l)=1+\sum_{n=1}^\infty \ \sum_{\sigma_1+\cdots+\sigma_{2n}=0}
\l^{2n}a_{n}(-\sigma_1,\ldots,-\sigma_{2n})
\ J^{(vac)}(\sigma_1,\ldots,\sigma_{2n}) \ ,
}
where the sum is taken over all sets of variables
$\sigma_1,\ldots,\sigma_{2n}=\pm1$ with zero total sum and 
$ J^{(vac)}(\sigma_1,\ldots,\sigma_{2n})$ denote vacuum eigenvalues 
of the operators  \Js. 
The numerical 
coefficients $a_n$ defined as 
\eqn\acoef{
a_{n}(\sigma_1,\ldots,\sigma_{2n})=q^n\, Z_+^{-1}(p)\,{\rm
Tr}_{\rho_+}\,(e^{2\pi i 
p{\cal H}}{\cal E}_{\sigma_1}\cdots {\cal E}_{\sigma_{2n}})\ ,}
where trace is taken over the representation $\rho_+$ of the $q$-oscillator 
algebra \qosc\ and $Z_+(p)$ is given by \zpm. 
It is easy to see that
\eqn\Jbound{
\sum_{\sigma_1+\cdots+\sigma_{2n}=0}\Big|
J^{(vac)}(\sigma_1,\ldots,\sigma_{2n})\Big|\le {\cal Z}_n\ .}
To estimate the coefficients \acoef\ it is convenient to use the explicit 
form of the 
representation matrices $\rho_+({\cal E}_\pm)$ and $\rho_+({\cal H})$
given in\ (D.6).   Using these one can show
\eqn\abound{
|a_n(\{\sigma\})|< {2^{2n}\over \Big|
\prod_{j=1}^{n}(1-q^{-2j}e^{4\pi i P}) 
\Big|}
}
where we have assumed that 
\eqn\restr{2p\not=n\beta^2+m} for any integer $m$ and
any positive integer $n$. For rational $\beta^2$ the relation \abound\
obviously imply 
\eqn\arat{|a_n(\{\sigma\})|<C^n}
is $C$ is some constant. Combining \arat, \Jbound\ and \Zbound\ we conclude 
the series \Avac\ in this case converges  in a whole complex plane of
$\l^2$.
In fact, the same inequality \arat\ 
holds for
(almost all) irrational $\beta^2$. This follows from a remarkable
result of \refs{\HL,\driver}
\eqn\hardy{
\lim_{n\rightarrow\infty} {1\over n} \log  \Big| 
\prod_{j=1}^{n}(1-q^{-2j}e^{4\pi i P})
\Big| =\int_0^{2\pi}\log(2\sin x)\,dx=0\ ,}
which is valid for all irrational $\beta^2$ satisfying \restr\ 
except a set of some exceptional irrationals of the linear Lebesgue 
measure zero (see \refs{\HL, \driver} for the details).

\appendix{C}{}
Using\ \vars,\ \tens,\ \nuyt\ one can write the Virasoro generator $L_0$ as
\eqn\lz{
L_0={P^2\over \beta^2}+{c-1\over24}+{2\over \beta^2}\ \sum_{n>0} a_{-n}a_n\ .
}
Then it easy to show that
\eqn\lphi{
[L_0,\varphi(u)]=-i\, \partial_{u}\varphi(u)\ .
}
Therefore the adjoint action of the the operator $\exp(i \veps
L_0)$ on \Tf\
\eqn\Adj{
e^{i \veps L_0} \ {\bf T}^{(f)}_j(\l)\ e^{-i \veps L_0}=
\Tr_{\pi_j}\left[e^{i (\pi P +f) H}
{\cal P} \exp\Big( \int_\veps^{2\pi+\veps} K(u) du \Big)\right]
}
reduces to a shift of the limits of integration in the ${\cal
P}$-exponent on the amount of $\veps$, where $\veps$ is assumed to be
real. Here $K(u)$ is the same as in \Kdef. Retaining in  \Adj\
linear in $\veps$ terms only one gets
\eqn\LTcom{
[L_0,{\bf T}^{(f)}_j(\l)]=-i\, \Tr_{\pi_j}\left[e^{i (\pi P +f) H}\Big(
K(2\pi)\,e^{-i \pi PH}\,{\bf L}_j(\l)-  e^{-i \pi PH}\,{\bf L}_j(\l)\,
K(0)\Big)\right]\ .
}
Expanding the ${\cal P}$-exponent as in \Lexpan, using \vper, the
commutations relations \vcomm\ and \uals\ and the cyclic property of
the trace one obtains
\eqn\LTcomm{
[L_0,{\bf T}^{(f)}_j(\l)]=\sin(\pi
P-f)\sum_{\s_0+\cdots+\s_n=0}a^{(f)}(\s_0,\s_1,\ldots,\s_n)
:e^{-2\s_0\varphi(2\pi)}:\, J(-\s_1,\ldots,-\s_n)\, ,
}
where
\eqn\af{
a^{(f)}(\s_0,\s_1,\ldots,\s_n)=-2\, \sigma_0\ e^{i\s_0(\pi P-f)} \Tr_{\pi_j}
\bigg[e^{i(\pi P+f)H}\, y_{\s_0}y_{\s_1}\cdots y_{\s_n}\bigg]
}
with
$$
y_+=\l\ q^{H/2}E\, ,\qquad y_-=\l\ q^{-H/2}F\, ,
$$
and the ordered integrals $ J(\s_1,\ldots,\s_n)$ defined in \Js.
Obviously, RHS of \LTcomm\ vanishes if
$$
f=\pi (P+N)\ ,
$$
where $N$ is arbitrary integer. We set $N=0$, since \Tf\ depends on $N$ 
only through a trivial sign factor $(-1)^{2 j N}$.  

Thus the operators ${\bf
T}_j(\lambda)$\ \Tj\ commute with the simplest local
IM ${\bf I}_1=L_0-c/24$.
As it follows from \funcrel\ and \Thf\ the coefficients of the series 
expansions of ${\bf
T}_j(\lambda)$\ in the variable $\l^2$ can be algebraically expressed 
in terms the nonlocal IM \Gn. Therefore the above commutativity
is equivalent to
\eqn\Gni{
[{\bf G}_n,{\bf I}_1]=0\, ,\qquad n=1,2,\ldots,\infty\ .}
In fact, the operators ${\bf G}_n$ commute 
with all local IM \loim.   To check this one has to
transform the ordered integrals in \Gn\ to contour integrals. 
For example,
${{\bf G}_1}$ can be written as \BLZZ\ 
\eqn\Gone{\eqalign{
&{ {\bf G}_1} =
\big(q^2-q^{-2}\big)^{-1}\  \int_{0}^{2\pi}d u_1 \int_{0}^{2\pi}
d u_2\
\bigg\{\  \big(q e^{-2\pi i P} - q^{-1} e^{2\pi i P}\big)\times
\cr &
V_{-}(u_1 + i0) V_{+}(u_2 - i0) +
\big(q e^{2\pi i P} - q^{-1} e^{-2\pi i P}\big)\ 
V_{+}(u_1 + i0) V_{-}(u_2 - i0)\ \bigg\} \ . }}
Characteristic property of the local IM is that their commutators 
with the exponential fields \eip\ reduces to a total derivative 
\refs{\Frenk,\FF}
\eqn\kuyt{[{\bf I}_{2n-1},V_\pm(u)]=\partial_u
\Big\{:O^{\pm}_n(u)V_\pm(u):\Big\}\equiv\partial_u X_n^\pm(u) .}
Here \ $O^{\pm}_n(u)$\ are  some  polynomials with respect the
field $\partial_u\varphi$ and its derivatives.
It follows then that the commutator of \Gone\ with ${\bf I}_{2n-1}$,
\eqn\Cn{{\bf C}_n=[{\bf I}_{2n-1},{\bf G}_1]\, ,}
is 
expressed as a double contour  integral of a linear combination 
of products of the form $\partial_{v_1}X_n^\pm(v_1)\,V_\mp(v_2)$ and  
$V_\pm(v_1)\,\partial_{v_2}X_n^\mp(v_2)$. It is important to note that 
the operator product expansion for these products does not contain
any terms proportional to negative integer powers of the difference
$(v_1-v_2)$. Therefore the above integrand for \Cn\ does not contain 
any contact terms (i.e. the terms proportional to the delta function 
$\delta(u_1-u_2)$ and its derivatives). Thus we can easily perform 
one integration 
\eqn\Cnn{\eqalign{
{\bf C}_n=&\big(q^2-q^{-2}\big)^{-1}\big(q e^{-2\pi i P} - q^{-1} e^{2\pi i
P}\big)\big(q e^{2\pi i P} - q^{-1} e^{-2\pi i P}\big)\times\cr
&\int_{0}^{2\pi}d u \
\, \bigg[
qe^{2\pi iP}\,V_-(u)X_+(0)- 
q^{-1} e^{-2\pi i P}X_-(0) V_{+}(u) + 
q e^{-2\pi i P}V_{+}(u)X_{-}(0)-\cr &\ \ \ \ \ \ \ \  q^{-1} e^{2\pi i
P}X_{+}(0)V_{-}(u)\ \bigg] \ ,}}
where we have used the periodicity property
\eqn\Xper{
X_\pm(u+2\pi)=q^{-2}\, e^{\pm4\pi iP}\, X_\pm(u)\ .
}
Using now the commutation relations
\eqn\XVcom{
X_\pm(0)\, V_\mp(u)=q^2\, V_\mp(u)\,  X_\pm(0)\, , \qquad u>0\ ,
}
where\ $\sigma_1,\sigma_2=\pm 1$, one can see that the RHS of
\Cnn\ is equal to zero. 
The higher nonlocal IM ${\bf G}_n$ also admit contour integral 
representations similar to \Gone\ and their commutativity with 
${\bf I}_{2k-1}$ can, in priciple, be proved in the same way. 
However, these representations become more and more complicated 
for high orders and in general unknown. 
It would be interesting to obtain a general proof of 
the above commutativity to all orders\foot{
B.Feigin and E.Frenkel
have pointed out \FFprivate\ that such proof 
can be obtained by extending the results of  \refs{\Frenk,\FF}.}.

\def\smatrix#1{\left[\matrix{#1}\right]}
\appendix{D}{}
In this Appendix we present the derivation of the factorization\ \factor.
Using the definition \Adef\ one can write the product of the operators 
${\bf A}_\pm$ from\ \factor\ in the form
\eqn\Aprod{
{\bf A}_+(\l\mu){\bf A}_-(\l\mu^{-1})=\big(Z_+(P)Z_-(P)\big)^{-1}\, 
{\rm Tr}_{\rho_+\otimes \rho_-}\Big[e^{i\pi P\overline{\cal H}}\,
{\bf L}_+(\l\mu)\otimes {\bf  L}_-(\l\mu^{-1})\Big]\ ,
}
where 
\eqn\mdef{\mu=q^{j+{1\over2}}\ ,}
and
the trace is taken over the direct product of the two representations
$\rho_+\otimes\rho_-$ of  \qosc\ (these are defined after \lmap\ in
the main text) and 
\eqn\Hbar{\overline{\cal H}={\cal H}\otimes 1- 1\otimes {\cal H}
}
It is convenient to choose the representation space of
$\rho_+$ ($\rho_-$) as a
highest module
generated by a free action of the operator $\rho_+[{\cal E}_-]$
$\big(\rho_-[{\cal E}_+]\big)$\ on a vacuum vector defined respectively as 
\eqn\vacv{
\rho_{\pm}[{\cal E}_\pm]\,  |0\rangle_\pm =0\, ,
\qquad \rho_{\pm}[ {\cal H}]\, |0\rangle_\pm
=0\ .
}
Defining  natural bases in these modules
\eqn\bas{
|k\rangle_\pm=\rho_{\pm}\big[
{\cal E}_\mp^k\big]\, |0\rangle_\pm\ ,\qquad k=0,1,2,\ldots,\infty\ ,
}
with the upper signs for $\rho_+$ and the lower signs for $\rho_-$ one can
easily calculate the matrix elements 
\eqn\matel{
\eqalign{
\rho_\pm[{\cal E}_\pm]\, |k\rangle_\pm={1-q^{\mp2k}\over(q-q^{-1})^2}\, 
|k-1\rangle_\pm\ ,&\qquad
\rho_\pm[{\cal E}_\mp]\, |k\rangle_\pm=|k+1\rangle_\pm\ ,\cr
\rho_\pm[{\cal H}]\, |k\rangle_\pm&=\mp2k\, |k\rangle_\pm\ .
}}
Notice that the trace in \zpm\ for this choice of $\rho_\pm$ reads
\eqn\Zpm{Z_+(P)=Z_-(P)={e^{2\pi i P}\over 2i \sin (2\pi P)}\ .}
Specializing now the formula \Lprod\ for the product of the two operators 
${\bf L}_\pm$ in \Aprod\ one obtains
\eqn\Lpr{
 {\bf L}_+(\l\mu)\otimes{\bf L}_-(\l\mu^{-1})=
e^{i\pi P \overline{{\cal H}}}\  {\cal P} \exp \bigg(\l
\int_0^{2\pi} \big(
V_-(u)\,q^{{\overline{{\cal H}}\over2}}\,\overline{{\cal E}}\ +
V_+(u)\,q^{-{\overline{{\cal H}}\over2}}\,\overline{{\cal F}}\,
\big) du \bigg)\ ,}
where $\overline{\cal H}$ is given by \Hbar\ and 
\eqn\EFbar{\eqalign{
\overline{{\cal E}}&=\mu\,{\cal E}_+\otimes q^{-{{\cal H}\over2}}+
                     \mu^{-1}\, q^{-{{\cal H}\over2}}\otimes {\cal E}_-\, ,\cr
\overline{{\cal F}}&=\mu \,{\cal E}_-\otimes q^{{\cal H}\over2}+
                     \mu^{-1}\,q^{{\cal H}\over2}\otimes {\cal E}_+\, .\cr
}}
The last two equations can be written in a compact form 
\eqn\EFab{
\overline{{\cal E}}=a_-+b_-\ ,\qquad \overline{{\cal F}}=a_++b_+\ ,}
if one introduces  the operators 
\eqn\abdef{
a_\pm=\mu\, {\cal E}_\mp\otimes q^{\pm{{\cal H}\over 2}}\ ,\qquad
b_\pm=\mu^{-1}\,  q^{\pm{{\cal H}\over 2}}\otimes{\cal E}_\pm\ ,
}
acting in $\rho_+\otimes \rho_-$. These operators satisfy the commutation
relations 
\eqn\abcom{\eqalign{
a_{\sigma_1}\, b_{\sigma_2}&=q^{2\sigma_1\sigma_2}\ 
b_{\sigma_2}\, a_{\sigma_1}\ ,\qquad [\overline{{\cal H}},a_\pm]=\mp2
a_\pm\ ,\qquad [\overline{{\cal H}},b_\pm]=\mp2
b_\pm\ ,\cr
q\,  a_-a_+&-q^{-1} a_+a_-={\mu^2\over q-q^{-1}}\ ,\qquad
q\,  b_+b_--q^{-1} b_-b_+={\mu^{-2}\over q-q^{-1}}\ ,\cr}}
where $\sigma_1,\sigma_2=\pm1$.

Further, the direct product of the modules $\rho_\pm$ can be decomposed
in the following way
\eqn\decom{
\rho_+\otimes \rho_-=\mathop{\oplus}_{m=0}^\infty\, \rho^{(m)}\ ,}
where each $\rho^{(m)}$, $m=0,1,2,\ldots,\infty$, is again a highest
weight module spanned on the vectors 
\eqn\Vmdef{
\rho^{(m)}: \qquad  |\rho^{(m)}_k\rangle=(a_++b_+)^k\, (a_+-\gamma b_+)^m
\  |0\rangle_+\otimes|0\rangle_-\ ,\qquad k=0,1,2,\ldots,\infty\ .}  
The constant $\gamma$ here is constrained by the relation
\eqn\gcon{
\gamma\not=-q^{-2n}\, ,\qquad n=0,1,2,\ldots,\infty\ ,}
but otherwise arbitrary.
To prove that the modules $\rho^{(m)}$ are linearly independent (as
subspaces in the vector space $\rho_+\otimes \rho_-$) it is
enough to prove that
$\ell+1$ vectors $|\rho^{(\ell-k)}_k\rangle$, $k=0,1,\ldots,\ell$, on each 
``level'' $\ell=0,1,\ldots,\infty$ are linearly independent (the
vectors on different levels are obviously linearly independent). To see
this let us use the commutation relations \abcom\ and rewrite  
\eqn\polyn{
z_k=(a_++b_+)^k\, (a_+-\gamma b_+)^{\ell-k},\qquad k=0,1,\ldots,\ell}
as ordered polynomials in the variables $a_+$ and $b_+$
\eqn\ordp{
z_k=\sum_{m=0}^\ell C^{(\ell)}_{km}\ \,  (a_+)^{\ell-m}\, (b_+)^m\
.}
If $\gamma$ satisfies \gcon\ the determinant of the coefficients of
these polynomials 
\eqn\cdet{
\det\big\Vert C^{(\ell)}_{km}\big\Vert_{0\le k,m\le \ell}=
\prod_{n=0}^{\ell-1} (\gamma+q^{-2n})^{\ell-n}}
does not vanish. That implies the required linear independence.

{}From the above definitions it is easy to see that the operators
$\overline{{\cal H}}$ and $\overline{{\cal F}}$ entering \Lpr\
act invariantly in each module $\rho^{(m)}$ 
\eqn\HFact{
\overline{{\cal H}},\ \overline{{\cal F}}\ :\qquad \rho^{(m)}\longrightarrow
\rho^{(m)}\ ,}
while for the operator $\overline{{\cal E}}$ acts as 
\eqn\Eact{
\overline{{\cal E}}\ :\qquad \rho^{(m)}\longrightarrow
\rho^{(m)}\oplus \rho^{(m-1)}}
with $\rho^{(-1)}\equiv0$.   The matrix element of these operators 
can be easily found from\ \EFab,\ \abcom,\
\Vmdef,
\eqn\efhmat{\eqalign{
(\rho_+\otimes\rho_-)[\overline{\cal H}]\,  |\rho^{(m)}_k\rangle
&=-2\, (m+k)\, |\rho^{(m)}_k\rangle,\cr
(\rho_+\otimes\rho_-)[\overline{\cal F}]\, | \rho^{(m)}_k\rangle&=
|\rho^{(m)}_{k+1}\rangle,\cr
(\rho_+\otimes\rho_-)[\overline{\cal E}]\,| \rho^{(m)}_k\rangle&
=[k]_q\,  [2j-1+k]_q \, | \rho^{(m)}_{k-1}\rangle+
c^{(m)}_k\, | \rho^{(m-1)}_k\rangle\ ,\cr}}
where we have used\ \mdef. The values of $c^{(m)}_k$ can be calculated 
but not necessary in that following. 
Thus the matrices \efhmat\ have the block triangular
form with an infinite number of diagonal blocks. It is essential to note that 
in each diagonal block 
these matrices coincide with those of the highest weight representation 
$\pi_j^+$ given by \spinj\  (up to an overall shift in the matrix 
elements of $(\rho_+\otimes\rho_-)[\overline{\cal H}]$ in different  blocks).
Substituting now \efhmat\ into \Lpr\ and then into
\Aprod\ and using the definition 
\Tjplus\ one easily arrives to the factorization \factor.

\listrefs
\end